\documentclass[fleqn,usenatbib]{mnras}
\usepackage{float}

\usepackage[T1]{fontenc}
\DeclareRobustCommand{\VAN}[3]{#2}
\let\VANthebibliography\thebibliography
\def\thebibliography{\DeclareRobustCommand{\VAN}[3]{##3}\VANthebibliography}

\usepackage{graphicx}	
\usepackage{amsmath}
\usepackage{multirow}
\usepackage{multicol}
\usepackage{color}
\usepackage[dvipsnames]{xcolor}

\usepackage{soul}

\usepackage{mathptmx}

\usepackage[T1]{fontenc}
\usepackage{ae,aecompl}

\usepackage{amssymb}	
\usepackage{longtable}
\usepackage{pdflscape}
\usepackage{anyfontsize}

\def\aj{AJ}%
\def\actaa{Acta Astron.}
\def\araa{ARA\&A}
\def\apj{ApJ}
\def\apjl{ApJ}
\def\apjs{ApJS}
\def\aap{A\&A}
\def\aaps{A\&AS}
\def\apss{Ap\&SS}
\def\mnras{MNRAS}
\def\pasp{PASP}
\newcommand{\tess}{{\it TESS}}
\newcommand{\kepler}{{\it Kepler}}
\newcommand{\ktwo}{{\it K2}}
\newcommand{\gaia}{{\it Gaia}}
\newcommand{\teff}{$T_\textrm{eff}$}
\newcommand{\logg}{$\log g$}
\newcommand{\feh}{[Fe/H]}
\newcommand{\mh}{[M/H]}
\newcommand{\mdot}{M$_\odot$}
\newcommand{\porb}{$P_{\rm orb}$}
\newcommand{\cd}{d$^{-1}$}
\newcommand{\logL}{$\log (L/L_{\odot})$}
\newcommand{\kms}{km\,s$^{-1}$}
\newcommand{\vsini}{$\upsilon\sin i$}
\newcommand{\vrot}{$v_{\rm rot}$}

\newcommand{\dsct}{$\delta$\,Sct}
\newcommand{\gdor}{$\gamma$\,Dor}

\def\spose#1{\hbox to 0pt{#1\hss}}
\def\lta{\mathrel{\spose{\lower 3pt\hbox{$\mathchar"218$}}
     \raise 2.0pt\hbox{$\mathchar"13C$}}}
\def\gta{\mathrel{\spose{\lower 3pt\hbox{$\mathchar"218$}}
     \raise 2.0pt\hbox{$\mathchar"13E$}}}

\title[Asteroseismology of HD\,23734, HD\,68703, and HD\,73345]{Asteroseismology of HD\,23734, HD\,68703, and HD\,73345 using \ktwo-\tess\ Space-based Photometry and High-resolution Spectroscopy}
\author[Joshi et. al.]{
{Santosh Joshi}$^{1}$\thanks{E-mail: santosh@aries.res.in},
{Athul Dileep}$^{1,2}$,
{Eugene Semenko}$^{3}$,
{Mrinmoy Sarkar}$^{1}$,
{Otto Trust}$^{4,5}$,
{Peter De Cat}$^{6}$,
\newauthor
{Patricia Lampens}$^{6}$,
{Marc-Antoine Dupret}$^{7}$,
{Surath C. Ghosh}$^{1}$,
{David Mkrtichian}$^{3}$,
{Mathijs Vanrespaille}$^{8}$,
\newauthor
{Sugyan Parida}$^{9}$,
{Abhay Pratap Yadav}$^{9}$,
{Pramod Kumar S.}$^{10}$,
{P. P. Goswami}$^{11}$,
{Muhammed Riyas}$^{12}$,
\newauthor
{Drisya Karinkuzhi}$^{12}$
\\
$^{1}$Aryabhatta Research Institute of Observational Sciences, Manora Peak, Nainital- 263002, India\\
$^{2}$Department of Applied Physics, M.J.P Rohilkhand University, Bareilly, Uttar Pradesh-243006, India\\
$^{3}$National Astronomical Research Institute of Thailand (NARIT), Chiang Mai 50180, Thailand \\
$^{4}$Department of Physics, Mbarara University of Science and Technology, P. O. Box 1410, Mbarara, Uganda\\
$^{5}$Department of Physics, Faculty of Science, Kyambogo University, P.O. Box 1, Kyambogo, Uganda\\
$^{6}$Royal Observatory of Belgium, Ringlaan 3, B-1180 Brussels, Belgium \\
$^{7}$Space Sciences, Technologies and Astrophysics Research (STAR) Institute, University of Liège, B-4000 Sart Tilman, Belgium\\
$^{8}$Institute of Astronomy, KU Leuven, Celestijnenlaan 200D, B-3001 Leuven, Belgium\\
$^{9}$Department of Physics and Astronomy, National Institute of Technology Rourkela, Sector 1, Rourkela - 769008, Odisha-India \\
$^{10}$Indian Institute of Astrophysics, Koramangala, Bangalore-560034, India\\
$^{11}$Dakshin Kamrup College, Gauhati University, Assam-781125-India \\
$^{12}$Department of Physics, University of Calicut, Thenhipalam, Malappuram 673635, India\\
\\
}
\date{ Accepted DD Mmmm YYYY / Received DD Mmmm YYYY  } 
 
 \pubyear{2024}
 
\begin{document}
 \label{fitrstpage}
 
 \maketitle 
 
\begin{abstract}

In this paper, we present a comprehensive study of three stars, HD\,23734, HD\,68703, and HD\,73345, which were previously observed as chemically peculiar candidates within the Nainital-Cape survey and reported as null results for the pulsational variability. Frequency analyses of \ktwo\ and \tess\ time-series photometric data reveal the co-existence of rotational modulation and pulsation. We use the spectrum synthesis technique to determine fundamental parameters and chemical composition, which shows that all the three stars are likely to be chemically normal. The evolutionary status of the target stars corresponds to the main-sequence phases and places them within the $\delta$ Scuti instability strip of the Hertzsprung–Russell diagram. The line profile variability is observed in all three stars, especially intriguing in HD\,68703 and a typical signature of the non-radial pulsation, demands further detailed examination. Using \tess\ photometry, we identified the radial modes of orders $n$=3 and 4 for HD\,23734, $n$=1, 3, and 4 for HD\,68703, and $n$=3,4 and 5 for HD\,73345. In addition to the presence of pulsation and rotation, HD\,73345 exhibits a steady increase in radial velocity that we interpret as the star being likely to be part of a long-period binary system. Finally, we propose an extended campaign aimed for the in-depth spectroscopic and spectropolarimetric study of selected pulsating stars monitored under the Nainital-Cape survey project. 

\end{abstract}




\section{Introduction}

About 10\% of upper main-sequence (MS) stars with spectral types ranging from B to F exhibit anomalous intensities of spectral features compared to normal stars of similar spectral types, ages, and populations. Owing to the observed photospheric overabundances of silicon (Si), chromium (Cr), strontium (Sr), europium (Eu), and many other heavy elements, and are classified as chemically peculiar or CP stars~\citep{2018Ghazaryan}.

The CP stars are traditionally divided into four subgroups based on their effective temperature, strength of absorption lines, and magnetic field properties: Am/Fm (CP1), Bp/Ap (CP2), HgMn (CP3), and He-weak (CP4) stars \citep{1974ARA&A..12..257P, 2021MNRAS.504.3758P}. As our study focuses primarily on Am (metallic-line A-type) and Ap (peculiar A-type) stars, we briefly introduce these two subtypes within the context of the present work.

The Am/Fm stars are non-magnetic objects characterised by deficiencies in light elements (e.g., C, N, O, Ca, Sc) and/or overabundances of Fe-peak elements (e.g., Fe, Cr) and rare-earth elements \citep{1970PASP...82..781C, 2019MNRAS.484.2530C}. These A- or early F-type MS stars occupy the same region of the Hertzsprung–Russell (H-R) diagram as chemically normal A-F stars, with effective temperatures in the range of 7000-9000\,K. Most of the Am stars are slow rotators with projected rotational velocities (\vsini) below 150\,\kms\ and are commonly found in binary systems with orbital periods (\porb) exceeding 1.2\,days \citep{2008CoSka..38..129I}.

The Ap stars possess strong magnetic fields and exhibit pronounced overabundances of elements such as Si, Sr, Cr, and Eu. The magnetic fields of Ap stars are generally organised as dipoles and can reach strengths of several tens of kilogauss \citep{2021A&AT...32..137B}, often varying with rotational period \citep{1947ApJ...105..105B, 2007A&A...475.1053A}. Although the origin of these fields remains uncertain, they are thought to be fossil in origin, retained from the process of star formation \citep{2004Natur.431..819B, 2007A&A...475.1053A}. 

The axial rotation of CP stars has a powerful impact on various stellar properties. The chemical peculiarities of Am and Ap stars are believed to arise from processes like atomic diffusion, which can work only in a stabilised environment. It is the slow rotation and magnetic field that suppress convective mixing and allow for the selective levitation or gravitational settling of chemical elements in stellar interiors \citep{1970ApJ...160..641M, 2000ApJ...529..338R}. Slow rotation as the main factor of emerging atomic diffusion in Am stars is typically associated with tidal synchronisation in binary systems, while in Ap stars it is attributed to strong magnetic fields \citep{1973ApJS...25..137A, 1974ApJ...191..165S, 2016MNRAS.460.1912G}.
Nonetheless, in certain instances, abnormally high rotational velocities have been noted \citep[e.g.,][]{2022A&A...668A.159M}. The rotational properties of a few hundred magnetic CP stars have recently been examined in open star clusters by \cite{2017MNRASNetopil}, and the authors found that the evolution of rotational periods is consistent with the conservation of angular momentum during the main sequence (MS) phase. The coexistence of pulsations, surface chemical inhomogeneities, moderate rotation, and magnetic fields makes Am and Ap stars exceptional laboratories for asteroseismic modelling, offering valuable insights into internal stellar structure and evolutionary processes \citep{2010A&A...511L...7M, 2013A&A...556A..18K}.

 Several ground-based surveys have been conducted in the past to study CP stars, including the Cape and Nainital-Cape Surveys \citep{1991MNRAS.250..666M, 2001AA...371.1048M}, the SuperWASP\footnote{SuperWASP : Super Wide Angle Search for Planets} Survey \citep{2011A&A...535A...3S}, ASAS-3\footnote{ASAS : All Sky Automated Survey} \citep{2015A&ABernhard}, ${\it 2MASS}$\footnote{2MASS : Two Micron All-Sky Survey} \citep{2016A&AHerdin}, LAMOST\footnote{LAMOST : Large Sky Area Multi-Object Fiber Spectroscopic Telescope} \citep{2020A&AHummerich, 2023A&A...676ALabadie}, ATLAS Sky Survey \citep{2021MNRASBernard}, and the ZTF\footnote{ZTF :  Zwicky Transient Facility} Survey \citep{2024A&AHummerich}, to name a few. Thanks to the high-precision, continuous photometric data obtained from space missions such as BRITE\footnote{BRITE : BRIght Target Explorer} \citep{2024bss..confEPaunzen}, {\kepler} \citep{2018A&AHummerich}, GAIA\footnote{GAIA : Global Astrometric Interferometer for Astrophysics} \citep{2024A&APaunzen}, and {\tess}\footnote{TESS : Transiting Exoplanet Survey Satellite} \citep{2020pase.conf..214K}, where the data acquired by these missions have led to the detection of  multi-periodic pulsations in numerous CP stars covered by the \dsct\ instability strip in the H-R diagram \citep{2017MNRAS.465....1S, 2021MNRAS.506.1073H, 2024A&A...690A.104D}, providing valuable data for modelling the pulsation \citep{2015Joshi, 2024Paunzen}.
The CP stars have also been searched for and studied in open star clusters by various authors, as they provide strong constraints on their ages and metallicity given that they are assumed to have originated from the same molecular clouds \citep{2025MNRASFaltov}. One of the intriguing findings on CP stars in open clusters is that they are of very young ages (log $t > 6.90$) and located at galactocentric distances up to 11.4\,kpc \citep{2005A&APaunzen, 2014A&APaunzen}.

A class of pulsating variables known as $\delta$ Scuti (\dsct) stars are located near the intersection between the classical instability strip and the main sequence. They span effective temperatures of 6300 $<$ \teff $<$ 9000\,K, luminosities of 0.6 $<$ \logL\ $<$ 2.0\,dex, and masses (M) between 1.5 and 2.5\,\mdot. These stars can be found in pre-main sequence stages or just evolving off the main sequence and exhibit low-order pressure ($p$) modes with pulsation periods ranging from $\sim$15\,minutes to $\sim$8\,hours \citep[e.g.,][]{2001A&A...366..178R, 2005ApJ...619.1072B, 2021FrASS...8...55G}. The primary excitation mechanisms include the $\kappa$-mechanism in the He\,{\sc ii} ionization zone \citep{ba01400g, 2022A&A...664A..32S} and turbulent pressure in the H ionization layer \citep{2014ApJ...796..118A}.
The \dsct\ pulsators represent a particularly intriguing class of variable stars due to their location in the transition region between stars with convective envelopes (M $<$ 2\,\mdot) and those with radiative envelopes (M $>$ 2\,\mdot). This transitional nature makes them ideal targets for probing the structure of stellar envelopes. However, understanding the mechanisms that drive pulsations and determining mode selection in \dsct\ stars remains a significant unresolved challenge in stellar astrophysics, as their masses fall within a regime where convective zones persist in the outer stellar layers. In binary systems, tidal interactions may further influence the pulsation patterns and rotation rates, producing observable effects in light curves and spectra.

The ``Nainital–Cape Survey'' (N-C Survey), a collaborative effort between the Aryabhatta Research Institute of Observational Sciences (ARIES) in India and the South African Astronomical Observatory (SAAO) in South Africa, was initiated in the late 1990s to search for short-period pulsations in CP stars belonging to the northern and southern hemispheres and led to the discovery and detailed study of the first far-northern roAp star, HD\,12098 \citep{2001A&AGirish, 2024MNRAS.529..556K}. Subsequent time-resolved photometric and spectroscopic observations revealed anomalous pulsations in half a dozen CP stars \citep{2003MNRAS.344..431J, 2006A&A...455..303J, 2009A&A...507.1763J, 2010MNRAS.401.1299J, 2012MNRAS.424.2002J, 2016A&A...590A.116J, 2017MNRAS.467..633J}. In the era of space-based photometry, \citet{2022MNRAS.510.5854J} discovered HD\,73619 as the first CP heartbeat star, lacking tidally induced pulsations. Recently, \citet{2024MNRAS.534.3211S} conducted an asteroseismic analysis of HD\,118660 using \tess\ light curves. A series of Non-Local Thermodynamic Equilibrium (NLTE) analyses of hot CP stars using data from HESP and HERMES spectrographs was also carried out by \citet{2020MNRAS.492.3143T, 2021MNRAS.504.5528T, 2023MNRAS.524.1044T}. Furthermore, \citet{2024BSRSL..93..227D} revisited the variability classification of N-C Survey stars using \tess\ data. To address the spatial resolution limitations of \tess, we have also initiated a ground-based search for pulsating CP stars in open clusters \citep{10.1093/mnras/staf361}. These efforts have now expanded into a global collaboration within the field of asteroseismology through various international bilaterals (e.g. {\it BINA \footnote{BINA : Belgo-Indian Network for Astronomy and Astrophysics}}) and multilaterals research cooperation (e.g. { \it SAPTARISI\footnote{ SAPTARISI : Search and Follow-up Studies of Time-domain Astronomical Sources using Sky Surveys, BRICS Telescopes and Artificial Intelligence}}).

In this study, we perform asteroseismic investigation of three multi-periodic $\delta$ Scuti stars, HD\,23734, HD\,68703, and HD\,73345,  using combined space-based photometric observations from the \ktwo\ and \tess\ missions, complemented by multi-epoch spectroscopy from ground-based facilities.  The manuscript is organised as follows. The criteria for the selection of the sample are given in Section~\ref{selection}. The Section~\ref{Photo} describes the photometric observations and frequency analysis, along with an overview of spectroscopic data acquisition and reduction. Section~\ref{spec_an} presents a detailed photometric and spectroscopic analysis to derive fundamental parameters and chemical abundances, along with the evolutionary status of the stars. In Section~\ref{modes}, we identify the observed radial pulsation modes, while Section~\ref{seismodel} provides details on seismic modelling to constrain the basic stellar parameters. Finally, Section~\ref{concl} summarises the conclusions drawn from the present study.

\section{Target Selection}
\label{selection}

 The N-C Survey comprises 381 stars, and none of them were observed by the \kepler\ space mission. However, \ktwo\ data are available for eight targets. Of these, five have been classified as rotational variables \citep{2022MNRAS.510.5854J}, and three stars -- HD\,23734, HD\,68703, and HD\,73345 -- have been identified as pulsating stars in the present study.
 
HD\,23734 was listed as a member of the Pleiades cluster based on Gaia DR2 and APOGEE radial velocity data \citep{2020ApJ...903...55P}. \citet{2006A&A...455..303J} considered this star as a CP candidate based on  Str\"omgren photometric indices. The age and metallicity of this cluster are reported as $\log(\mathrm{Age}) = 8.920$ and [Fe/H] = $-0.12\pm0.06$ \citep{2021MNRASDias}. However, using the Gaia DR3 data, \citet{2023A&A...677A.163A} excluded HD\,23734 from membership of the Pleiades.

HD\,68703 is included in the General Catalogue of Ap and Am stars \citep{2009A&A...498..961R}, where it is listed as an Am star \citep{1978AJ.....83..606M} with the spectral type A8-dD (with "dD" denoting a $\delta$ Delphini-type star \citep{1971PASP...83..296C}). Its metallicity, derived from LAMOST medium-resolution spectroscopy is \feh~= 0.352 \citep{2020ApJS..251...15Z}. Later, HD\,68703 has been classified as a non-magnetic star by \citet{2006MNRAS.372.1804K}, with a spectral type of F1IV \citep{1995ApJS...99..135A}, and it exhibits $\rho$ Puppis-type peculiarity — characteristic of evolved, pulsating Am stars. Fundamental parameter derived by \cite{2024A&A...690A.104D} are: \teff~= 7192\,K, \logg~= 3.8\,dex, and \logL~= 1.29. The same authors investigated its {\tess} light curves and detected five radial pulsation modes, although these remain unidentified.

HD\,73345, also known as CY Cnc, is a \dsct\ star with a reported pulsation period of 0.1\,d \citep{1990A&AS...83...51L} while based on the time-series analysis of the \tess\ data, \citet{2023A&APamos} reported 26 frequencies. This is a confirmed member of the Praesepe cluster, which has an estimated age of 759 million years \citep{2023A&APamos}. HD\,73345 was identified as a candidate CP star in the catalogue of \citet{1998A&AS..129..431H}. \citet{2002AJ....124..989G} reported this star as a mildly-peculiar star with slightly strong Ca II K. However,  \citet{2008A&A...483..891F} listed the fundamental parameters of the star as \logL~= 1.17, \teff~= 7993\,K, \logg~= 3.96\,dex and \feh~= 0.26, their abundance analysis concluded that the star is a chemically normal A7V spectral type star.

\section{Observations and Data Reduction} \label{Photo}

\subsection{\ktwo\ and \tess\ Photometry} \label{space}

The \ktwo\ (Kepler two-wheel; February 4, 2014 -- September 26, 2018) mission surveyed a $105^{\circ} \times 105^{\circ}$ region along the ecliptic plane by observing each field for approximately 80 days across 20 different pointings \citep{2014PASP..126..398H}. The mission provided two data types: short-cadence (1-minute) and long-cadence (30-minute) photometry, with an angular resolution of 4\,arcsec  ~per pixel, thereby reducing  contamination from nearby stars.

The \tess\, \citep{2015JATIS...1a4003R}, launched in 2018, is an all-sky photometric survey operating primarily in the Cousins I band. It covers 90\% of the sky, originally excluding a narrow band along the ecliptic. During its primary mission, \tess\ observed 26 sectors -- each spanning $24^{\circ} \times 96^{\circ}$ -- dividing each hemisphere into 13 sectors and monitoring them sequentially over one year, with each sector observed for approximately 27.4 days. With an angular resolution of 21\,arcsec~per pixel, \tess\ data may suffer from spatial blending and contamination from nearby stars. Sector overlaps at high ecliptic latitudes enable the repeated observation of some stars across multiple sectors. The primary data products include short-cadence (2-minute) observations for pre-selected targets and full-frame images (FFIs) captured every 30 minutes for all stars within the field of view. The extended \tess\ mission, initiated on July 4, 2020, introduced higher temporal resolutions, including 20-second, 200-second, and 10-minute cadences, allowing for a more detailed re-observation of all sectors.

\begin{table}
\centering
\caption{Summary of the observational details for the target stars observed by the \ktwo\ and \tess\ space missions, including the available campaigns/sectors and the respective cadences for each star.}
\label{obs_log}
\begin{tabular}{|c|c|c|c|}
\hline\hline
Star      & Mission & Campaigns/ & Cadence \\
          &         & Sectors    & (s)     \\
\hline
HD\,23734 & \ktwo   & 4          & 1800    \\
          & \tess   & 42, 43, 44 & 120     \\
          & \tess   & 70, 71     & 200     \\
HD\,68703 & \ktwo   & 5, 18      & 1800    \\
          & \tess   & 44, 45, 46 & 120     \\
          & \tess   & 71, 72     & 200     \\
HD\,73345 & \ktwo   & 5, 18      & 1800    \\
          & \tess   & 44, 45, 46 & 120     \\
          & \tess   & 72         & 200     \\
\hline\hline
\end{tabular}
\end{table}

The details of the data available from different \tess\ sectors and \ktwo\ campaigns are presented in Table~\ref{obs_log}. The photometric data products were retrieved using the Python package \textsc{lightkurve} \citep{2018ascl.soft12013L}, sourced from the Barbara A. Mikulski Archive for Space Telescopes (MAST)\footnote{\url{https://archive.stsci.edu/}}. For both the missions, we utilised the Pre-search Data Conditioning Simple Aperture Photometry (PDCSAP) flux, which has been corrected for instrumental systematics using Co-trending Basis Vectors (CBVs). Outliers were identified and removed through a combination of visual inspection and sigma clipping. The resulting light curves were then converted to relative milli-magnitudes (mmag) and normalised by subtracting the mean magnitude level to centre them around zero.

\begin{figure*}
\centering
 \includegraphics[width=\textwidth]{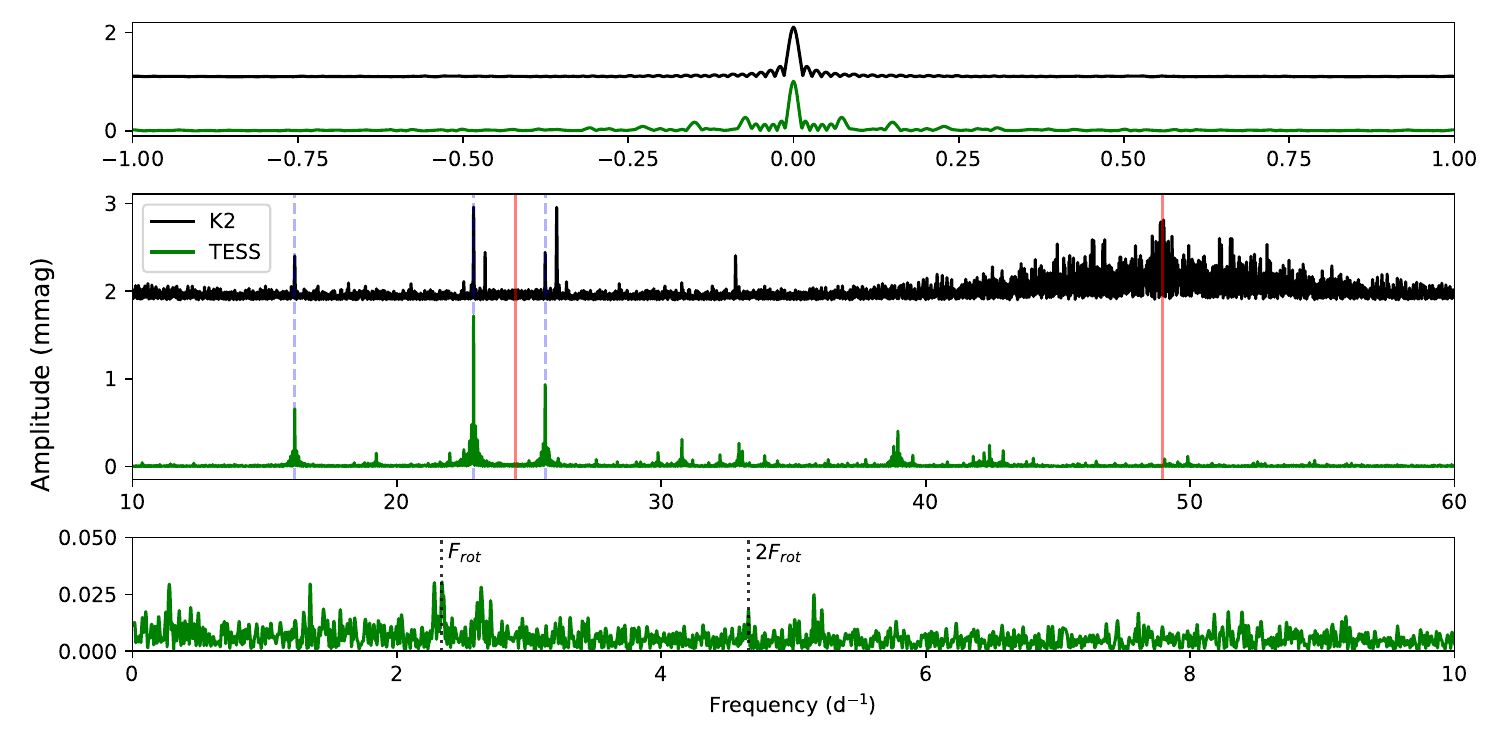}
 \caption{
\textit{Top panel:} The window function for HD\,23734 based on photometric data from \ktwo\ (Campaign 4; black) and \tess\ (Sectors 42, 43, and 44; green).
\textit{Middle panel:} Comparison of frequency spectra derived from \ktwo\ and \tess\ light curves. For visual clarity, the \ktwo\ spectrum has been vertically shifted by 1.8 mmag. Red vertical solid lines mark the integer multiples of the \ktwo\ Nyquist frequency, while blue dashed lines indicate frequencies common to both missions.
\textit{Bottom panel:} Zoomed-in view of the low-frequency range. The black dotted vertical lines denote the rotational frequency and its harmonics.
 }
\label{space_23734}
\end{figure*}

\begin{figure*}
\centering
 \includegraphics[width=\textwidth]{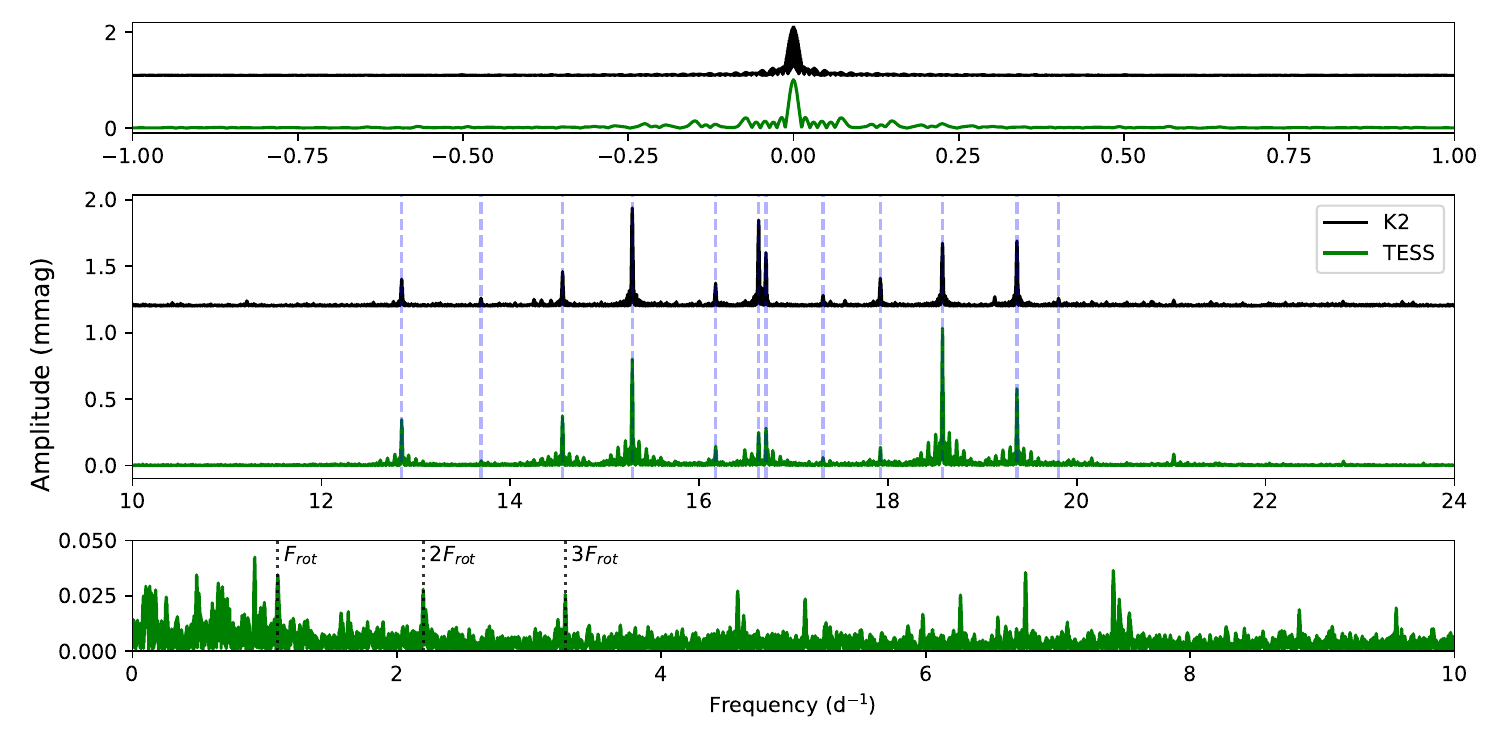}
 \caption{
Same as Fig.~\ref{space_23734}, but for HD\,68703 using \tess\ data from Sectors 44, 45, and 46, and \ktwo\ data from Campaigns 5 and 18. All detected frequencies lie within the Nyquist limits of both \ktwo\ and \tess.
 }
\label{space_68703}
\end{figure*}

\begin{figure*}
\centering
\includegraphics[width=\textwidth]{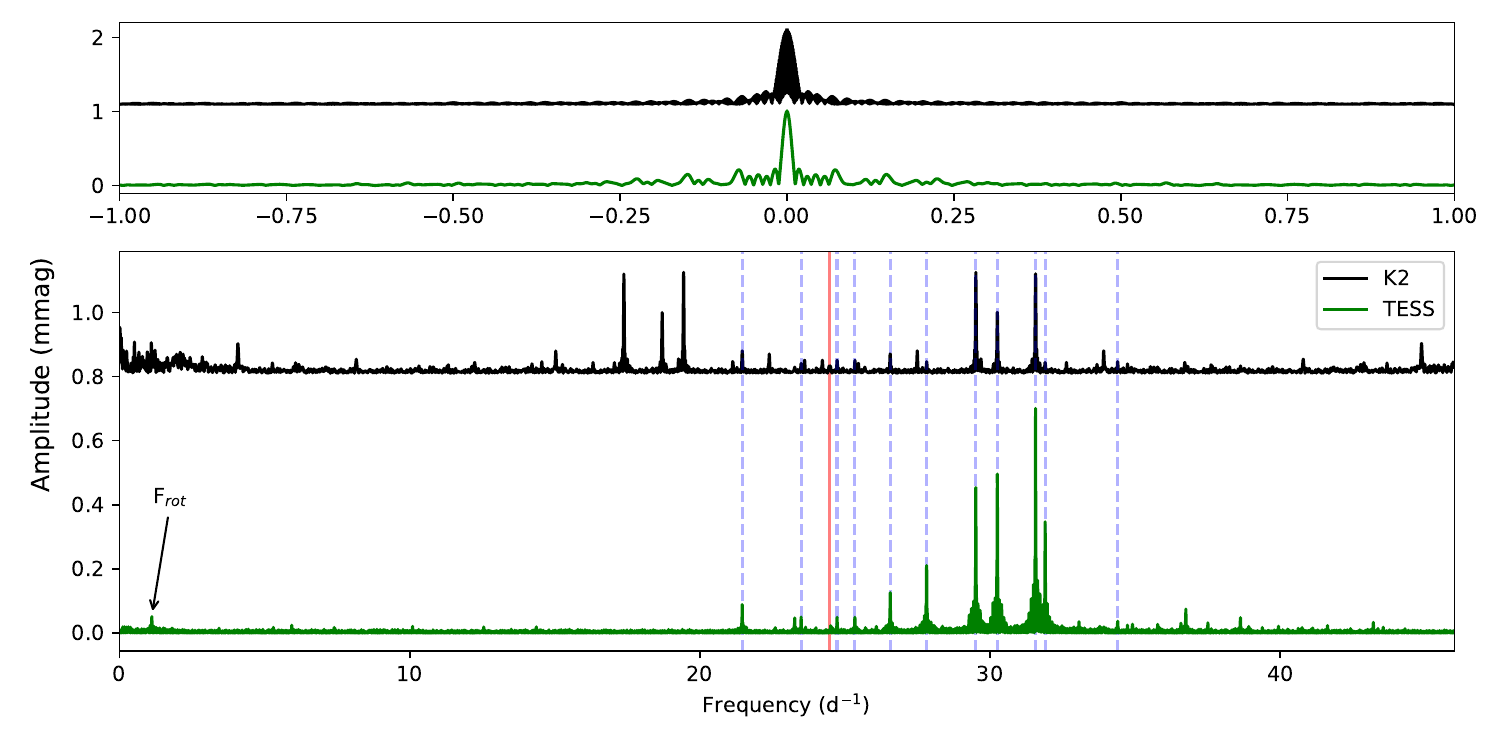}
\caption{
Same as Fig.~\ref{space_23734}, but for HD\,73345 using \tess\ data of Sectors 44, 45, and 46, and \ktwo\ data from Campaigns 5 and 18. The probable rotational frequency is indicated with an arrow.
}
\label{space_73345}
\end{figure*}

\subsubsection{Frequency Analyses}

The frequency analysis of the light curves was performed using the \texttt{Period04} programme \citep{2005CoAst.146...53L}. Periodograms were computed over a frequency range from zero to the Nyquist frequency, which is considered to be the maximum frequency resolvable in a frequency spectrum and defined as $\nu_{\mathrm{Nq}} = \frac{1}{2 \Delta t}$, where $\Delta t$ represents the time sampling interval. For long ($\Delta t = 29.43$\,min) and short ($\Delta t = 120$\,sec) cadence data, the Nyquist frequencies are approximately 24.27\,\cd\ and 360\,\cd, respectively. To minimise aliasing, we combined contiguous sectors while avoiding large gaps in the time-series data.
The resulting frequency spectra were iteratively prewhitened by fitting the frequency, amplitude, and phase of the highest peak using a sinusoidal model. Each identified frequency was successively subtracted from the light curve, and this process was repeated until the signal-to-noise ratio (SNR) of the residual peaks fell below the significance thresholds, namely SNR $<$ 5.2 for \tess\ data \citep{2021AcA....71..113B}, and SNR $<$ 5.7 for \ktwo\ data \citep{2015MNRAS.448L..16B}. Frequency and amplitude uncertainties were estimated using the formalism described by \citet{1999DSSN...13...28M}. Frequencies separated by less than 2.5 times the Rayleigh resolution ($1/\Delta T$) were considered unresolved \citep{1978Ap&SS..56..285L}.

Combined frequency spectra of HD\,23734, HD\,68703, and HD\,73345 -- based on \ktwo\ campaigns and \tess\ sectors -- are presented in Figs.~\ref{space_23734}, \ref{space_68703}, and \ref{space_73345}, respectively. Individual light curves and periodograms for each \ktwo\ campaign and \tess\ sector are shown in Figs.~\ref{fresp_k2_23734}--\ref{fresp_tess_73345} (left and right panels, respectively). A total of 64, 38, and 26 frequencies were recovered in HD\,23734, HD\,68703, and HD\,73345, respectively.
It is evident that the short-cadence \tess\ data enabled the detection of more frequencies compared to \ktwo. Some of the frequencies were found in one dataset (\ktwo) but absent in the other (\tess) -- or vice versa -- likely due to the wavelength dependency of pulsation amplitudes. The \ktwo\ band is centred around 6400 \AA\, while \tess\ operates near 9000 \AA. A comparison of the individual amplitude spectra of different \ktwo\ campaigns and \tess\ sectors also revealed amplitude modulations in certain frequencies, potentially caused by beating between close frequencies, non-linear mode coupling, or other non-linear effects \citep{2016MNRAS.460.1970B}.

We further searched for possible linear combinations involving radial and rotational frequencies, adopting a relation of the form $aF_n \pm bF_m$, where $a$ and $b$ are integers between $-5$ and $+5$, and $n$ and $m$ denote the order of the parent radial modes ($1, 2, 3, 4, ...$) or rotational frequency (`${\rm rot}$'). The identified frequencies, amplitudes, SNR, and their combinations for each target are summarised in Tables\,\ref{tab:tablet1}, \ref{tab:tablet2}, and \ref{tab:tablet3}.

\subsection{Spectroscopy} \label{specs}

For this study, we employed spectroscopic data obtained from the dedicated observing programmes and various archival sources. High-resolution spectra were acquired using the Hanle Echelle Spectrograph (HESP; \citealt{2015ExA....39..423C, 2018SPIE10702E..6KS}) of the 2.01-m Himalayan Chandra Telescope (HCT) at the Indian Astronomical Observatory (Ladakh, India) and with the High Efficiency and Resolution Mercator Échelle Spectrograph (HERMES; \citealt{2007ASPC..376..653D, 2011A&A...526A..69R}) working with the 1.2-m Mercator Telescope (La Palma, Spain). Medium-resolution spectra were collected for all programme stars with the Medium-Resolution Échelle Spectrograph (MRES) fed by fibre from the 2.4-m Thai National Telescope (TNT) at Doi Inthanon (Chiang Mai, Thailand). One of the programme stars, HD\,68703, was observed with MRES and HERMES in a time-series regime.

The raw spectra were reduced using dedicated pipelines specific to each instrument. For HESP and HERMES, those were their respective official pipelines. For the MRES data, we used the latest version of the Python-based pipeline PyYAP\footnote{Python-based Yet Another Pipeline (PyYAP): \url{https://github.com/ich-heisse-eugene/PyYAP}}.

Wavelength calibration for the medium- and high-resolution spectra was performed using Thorium-Argon (Th-Ar) reference sources, followed by correction for barycentric motion. All spectra were normalised to the continuum level using the \textsc{continuum} task in the Image Reduction and Analysis Facility (IRAF)\footnote{\url{https://iraf-community.github.io/}}. The signal-to-noise ratio (SNR) was calculated at a reference wavelength of 5500\,\AA.


The archival data were found in the European Southern Observatory (ESO) Science Portal\footnote{\url{https://archive.eso.org/scienceportal/home}}, a PolarBase archive of high-resolution spectropolarimetric stellar observations \footnote{\url{https://www.polarbase.ovgso.fr/}} \citep{2014PASP..126..469P}, and in the Data Release 10 (DR10) of the LAMOST archive\footnote{\url{https://www.lamost.org/dr10/v2.0/search}} \citep{2012RAA.Cui, 2012RAA.Zhao}.
Most of these data pertained to HD\,68703. This star was observed once with a multi-wavelength medium-resolution spectrograph XSHOOTER of the Very Large Telescope (VLT) of ESO under a project ID \textit{110.248M.001}. Two observations of HD\,68703 resulted in a total of 16 archival spectra from NARVAL, a high-resolution spectropolarimeter of the 2-m Bernard Lyot Telescope (TBL) at the Pic du Midi Observatory, and were acquired from PolarBase. All of our target stars were observed multiple times with the LAMOST, and their spectra were publicly released as part of DR10.

The details of all spectroscopic observations are summarised in Table~\ref{tab:speclog}. Besides the information about the origin and time of the observation, the last three columns of the table list the signal-to-noise ratio, radial velocity (RV), and projected rotational velocity (\vsini) measured from each spectrum.

\begin{table*}
\centering
\caption{
Overview of the spectroscopic observations acquired from various ground-based facilities. Asterisks denote the epochs when the time-series spectra were obtained; the listed parameters for those entries represent the values averaged across all individual exposures.
}  
\label{tab:speclog}
\fontsize{7.5}{9.0}\selectfont
\begin{tabular}{ccccccccccc}
\hline\hline
Object    & Telescope & Diameter & Spectrograph & Date & BJD        & Resolution                & Integration & SNR          & RV     & \vsini        \\
          &           & (m)      &              & (UT) & (2400000+) & ($\lambda/\Delta\lambda$) & Time (sec)  & (@5500\,{\AA}) & (\kms) & ($\pm$1 \kms) \\
\hline
HD\,23734 & LAMOST   & 4.0       & LRS          & 16 Jan 2018      & 58135.0181 &  1800 & 1800 & 140   & $-1.3\pm3.1$   &   - \\ 
          &          &           &              & 06 Jan 2019      & 58490.0631 &  1800 & 1800 & 227   & $-5.8\pm1.8$   &   - \\
          & HCT      & 2.01      & HESP         & 06 Dec 2018      & 58459.3276 & 30000 & 1200 & 137   & $-4.8\pm0.4$   & 133 \\
          &          &           &              & 06 Nov 2023      & 60255.1806 & 30000 & 1200 &  40   & $-4.6\pm0.9$   & 134 \\
          &          &           &              & 29 Jan 2024      & 60339.2102 & 30000 &  900 &  75   & $-4.4\pm0.7$   & 135 \\
          & Mercator & 1.20      & HERMES       & 04 Nov 2023      & 60252.5708 & 85000 &  200 &  60   & $-3.4\pm0.5$   & 132 \\
          & TNT      & 2.4       & MRES         & 13 Feb 2025      & 60720.0171 & 18500 &  900 & 170   &  $4.0\pm1.8$   & 138 \\
\hline
HD\,68703 & VLT UT3  & 8.2       & XSHOOTER     & 22 Dec 2022      & 59935.8470 & 11333 &   10 & 366   &  $0.0\pm4.5$   &   - \\
          & LAMOST   & 4.0       & MRS          & 27 Dec 2017      & 58115.2472 &  7500 &  600 & 164   &            -   &   - \\
          &          &           &              & 28 Nov 2018      & 58451.3479 &  7500 & 1200 &  21   &            -   &   - \\
          &          &           &              & 19 Dec 2018      & 58472.2812 &  7500 & 1200 &  21   &            -   &   - \\
          &          &           &              & 27 Dec 2018      & 58480.3013 &  7500 & 1200 &  41   &            -   &   - \\
          &          &           &              & 11 Feb 2019      & 58526.1180 &  7500 & 1200 &  11   &            -   &   - \\
          &          &           &              & 04 Jan 2021      & 59219.2638 &  7500 & 1200 &  79   &            -   &   - \\
          & HCT      & 2.01      & HESP         & 29 Nov 2018      & 58452.4225 & 30000 & 1200 & 103   & $-0.3\pm0.2$   &  69 \\
          &          &           &              & 26 Dec 2023      & 60305.2285 & 30000 &  900 & 222   &  $0.2\pm0.2$   &  68 \\
          &          &           &              & 28 Jan 2024      & 60338.4145 & 30000 &  900 & 162   &  $0.6\pm0.2$   &  65 \\
          &          &           &              & 07 Mar 2025      & 60742.1224 & 30000 &  900 & 146   &  $0.8\pm0.2$   &  68 \\
          &          &           &              & 07 Mar 2025      & 60742.3348 & 30000 &  900 & 168   &  $0.7\pm0.2$   &  68 \\
          & TNT      & 2.4       & MRES         & 18 Dec 2024$^{*}$ & 60663.2273 & 18500 &  410 & 200   &  $2.3\pm0.2$   &  69 \\
          &          &           &              & 19 Dec 2024$^{*}$ & 60664.4384 & 18500 &  389 & 222   &  $1.0\pm0.2$   &  68 \\
          &          &           &              & 13 Feb 2025      & 60720.0672 & 18500 &  420 & 200   &  $2.6\pm1.1$   &  67 \\
          & TBL      & 2.0       & NARVAL       & 12 Jan 2007$^{*}$ & 54113.6120 & 65000 &  600 & 266   & $1.46\pm0.05$  &  68 \\
          &          &           &              & 17 Jan 2007$^{*}$ & 54118.6024 & 65000 &  600 & 296   & $1.45\pm 0.07$ &  68 \\
          & Mercator & 1.20      & HERMES       & 15 Mar 2025      & 60750.5739 & 85000 &  300 & 148   &  $0.9\pm0.2$   &  68 \\
          &          &           &              & 15 Mar 2025      & 60750.5819 & 85000 &  300 & 173   &  $1.3\pm0.2$   &  68 \\
          &          &           &              & 16 Mar 2025      & 60751.4649 & 85000 &  300 & 151   &  $1.0\pm0.2$   &  68 \\
          &          &           &              & 16 Mar 2025      & 60751.5914 & 85000 &  300 & 134   &  $0.9\pm0.2$   &  67 \\
          &          &           &              & 17 Mar 2025      & 60752.4550 & 85000 &  300 & 183   &  $1.7\pm0.2$   &  67 \\
          &          &           &              & 17 Mar 2025      & 60752.5762 & 85000 &  300 & 159   &  $1.1\pm0.1$   &  66 \\
          &          &           &              & 18 Mar 2025      & 60753.5355 & 85000 &  300 & 161   &  $1.2\pm0.1$   &  65 \\
          &          &           &              & 20 Mar 2025$^{*}$ & 60755.5388 & 85000 &  180 & 140   & $1.42\pm0.02$  &  66 \\
          &          &           &              & 23 Mar 2025      & 60758.5722 & 85000 &  360 & 119   &  $1.5\pm0.1$   &  66 \\
\hline
HD\,73345 & LAMOST   & 4.0       & LRS          & 04 Jan 2016      & 57392.2045 &  1800 & 1800 & 312   & $24.9\pm2.1$   &  -  \\
          &          &           &              & 27 Jan 2016      & 57415.1631 &  1800 & 2100 & 140   & $28.4\pm3.3$   &  -  \\
          &          &           &              & 24 Feb 2016      & 57443.1606 &  1800 & 1800 & 255   & $29.8\pm2.2$   &  -  \\
          & HCT      & 2.01      & HESP         & 06 Dec 2018      & 58459.3733 & 30000 & 1200 &  79   & $32.5\pm0.2$   &  90 \\
          &          &           &              & 26 Dec 2023      & 60305.2426 & 30000 &  900 &  55   & $32.4\pm0.4$   &  89 \\
          &          &           &              & 28 Jan 2024      & 60338.4302 & 30000 &  900 &  63   & $33.0\pm0.3$   &  89 \\
          & TNT      & 2.4       & MRES         & 13 Feb 2025      & 60720.0945 & 18500 &  900 & 160   & $37.1\pm1.2$   &  90 \\
\hline\hline
\end{tabular}
\begin{minipage}{0.9\linewidth}
\end{minipage}
\end{table*}

\section{Stellar Fundamental Parameters and Evolutionary Status} \label{spec_an}

The reliable values of the fundamental stellar parameters, such as effective temperature (\teff), surface gravity (\logg), metallicity (\mh), and rotational characteristics in the form of the projected rotational velocity (\vsini) or period of axial rotation, are essential for any theoretical modelling. These parameters naturally restrict the range of possible solutions. The following subsections, review the observational approach to the determination of stellar parameters and its outcome for all the target stars.

\subsection{Interstellar Extinction} \label{ext}

The interstellar extinction alters the observed characteristics of the stars. High-resolution spectroscopy of stars in close proximity to the Sun, like objects in our study, is less sensitive to this effect. However, we cannot ignore the reddening when it comes to photometry, which we used for the preliminary evaluation of stellar parameters.

To account for the potential effect of reddening, we first examined the three-dimensional Bayestar dust maps \citep{bayestar, 2019arXiv190502734G}. Using \gaia\ parallaxes \citep{2018yCat.1345....0G} and Galactic coordinates from the SIMBAD database\footnote{\url{https://simbad.u-strasbg.fr/simbad/}} \citep{2000A&AS..143....9W}, the Bayestar models yielded zero reddening values for all targets in our sample.
However, more recent interstellar extinction maps given by \citet{2022A&A...661A.147L} and \citet{2022A&A...664A.174V} indicate non-zero cumulative absorption values ($A_{\rm 5500}$, at $\lambda$ = 5500 \AA) for all three stars. Since $A_{\rm 5500}$ closely approximates the commonly used extinction parameter $A_\mathrm{V}$, we accounted for these values in the analysis. Following them, the highest extinction $A_{\rm 5500} = 0.086$\,mag is observed for HD\,23734, while HD\,68703 and HD\,73345 exhibit $A_{\rm 5500}$ equals to 0.025 and 0.039\,mag, respectively. We adopted the values of $A_{\rm 5500}$ as the upper limit of interstellar reddening $A_\mathrm{V}$ and used them in the calculation of stellar luminosities listed in Table~\ref{tab:ubvbeta}.

A more direct method of estimating reddening involves measuring the equivalent widths of interstellar Na\,\textsc{i} D1 and D2 lines in high-resolution spectra. Among the three targets, interstellar features near the Na\,\textsc{i} D1 and D2 lines were securely detected only in HD\,23734. The equivalent widths of these features were measured as $W_{\lambda}$(D1) = 100\,m\AA\ and $W_{\lambda}$(D2) = 85\,m\AA. Applying the empirical calibration given by \citet{2012ebv}, we derived the colour excess of $E_{B-V}$, which corresponds to a total extinction of $A_\mathrm{V} = 0.06$--0.10\,mag (assuming $A_\mathrm{V} = 3.11 \times$ $E_{B-V}$), in a good agreement with the $A_{\rm 5500}$ value for this star. There was no measurable interstellar Na\,\textsc{i} absorption detected in the spectra of HD\,68703 or HD\,73345.

\subsection{Fundamental parameters and chemical composition}

\subsubsection{Photometry based evaluation}

Before proceeding with spectroscopic analysis, we used Geneva \citep{1980VA.....24..141G, 1988A&A...206..357R}, $uvby\beta$ \citep{1963QJRAS...4....8S, 1966ARA&A...4..433S, 1966AJ.....71..114C, 1998A&AS..129..431H}, and 2MASS \citep{2003yCat.2246....0C} photometry to derive stellar fundamental parameters. By combining colour indices from multiple photometric systems, we aimed to minimise discrepancies arising from potential binarity, which can significantly affect the total flux of the system. Multiple \teff\ values derived from different indices were averaged, and the results are presented in Table~\ref{tab:ubvbeta}, alongside the parameters derived from the other two photometric systems.

In our sample, Geneva photometric indices are available only for HD\,68703 and HD\,73345 \citep{2022A&A...661A..89P}. Applying the calibrations by \citet{1997A&AS..122...51K} to the $B2-V1$, $d$, and $m2$ indices, we obtained \teff\ = 7180\,K and \logg\ = 4.02\,dex for HD\,68703, and \teff\ = 7980\,K and \logg\ = 4.50\,dex for HD\,73345. For the metallicity \mh, this calibration yields values in the range of  $+$0.20 to $+$0.35\,dex for both stars. Notably, HD\,73345 exhibits an unusually high surface gravity.

Photometric indices of the $uvby\beta$ system are available for all three stars, as compiled by \citet{2015A&A...580A..23P}. To analyse these, we employed the calibrations developed by \citet{1985MNRAS.217..305M} and refined by \citet{1993A&A...268..653N}, providing estimates of \teff, \logg, and \mh\ for the stars under study.

From the 2MASS photometric system, we estimated stellar effective temperatures using the relations given by \citet{2021MNRAS.507.2684C}. The photometrically-derived parameters were subsequently refined using the spectra.

\begin{table*}
\centering
\caption{Basic physical parameters of the target stars retrieved from the SIMBAD database (Columns 2--4). The rest of the columns list parameters estimated from photometric calibrations. The distance $d$ is calculated using the \gaia\ DR3 parallax \citep{2023A&A...674A...1G}. Mean photometric values for each star are indicated with an asterisk.}
\label{tab:ubvbeta}
\begin{tabular}{ccccccccccc}
\hline
\hline
Star &$\alpha_{\rm J2000}$ & $\delta_{\rm J2000}$ & $V$             & $A_{V}$           & $M_{v}$        &$d$         & $T_{\rm eff}$   & \logg            & \mh                         \\
     & ($^h\;^m\;^s$)    & ($^{\circ}\;'\;''$) & ($\pm$0.01 mag) & ($\pm$0.001 mag) &($\pm$0.01 mag) &($\pm$1 pc) & ($\pm250$\,K) & ($\pm0.20$\,dex) & ($\pm0.20$\,dex)  \\
\hline
HD\,23734 & 03 48 10.59 & +21 19 44.66 & 7.96 & 0.086 & 2.56 & $120$ & 7623$^{\rm 1}$ & 4.09$^{\rm 1}$ & 0.12$^{\rm 1}$  \\
          &             &              &      &       &      &       & 7700$^{\rm 2}$ &   -          &   -                \\
          &             &              &      &       &      &       & 7660$^{*}$    & 4.09$^{*}$    & 0.12$^{*}$          \\
HD\,68703 & 08 14 11.14 & +17 40 33.43 & 6.47 & 0.025 & 1.55 &  $96$ & 7340$^{\rm 1}$ & 3.86$^{\rm 1}$ & 0.32$^{\rm 1}$  \\
          &             &              &      &       &      &       & 7180$^{\rm 2}$ &   -          &  -                \\
          &             &              &      &       &      &       & 7180$^{\rm 3}$ & 4.02$^{\rm 3}$ & 0.32$^{\rm 3}$       \\
          &             &              &      &       &      &       & 7230$^{*}$    & 3.93$^{*}$    & 0.32$^{*}$         \\
HD\,73345 & 08 38 37.86 & +19 59 23.09 & 8.14 & 0.039 & 1.82 & $184$ & 7710$^{\rm 1}$ & 3.99$^{\rm 1}$ & 0.26$^{\rm 1}$  \\
          &             &              &      &       &      &       & 7300$^{\rm 2}$ &   -          &   -                \\
          &             &              &      &       &      &       & 7980$^{\rm 3}$ & 4.49$^{\rm 3}$ & 0.25$^{\rm 3}$      \\
          &             &              &      &       &      &       & 7660$^{*}$    & 4.24$^{*}$    & 0.26$^{*}$         \\
\hline\hline
\end{tabular}
\begin{minipage}{0.9\linewidth}
$^{\rm 1}$ $uvby\beta$.
$^{\rm 2}$ 2MASS.
$^{\rm 3}$ Geneva.
\end{minipage}
\end{table*}

\subsubsection{Spectroscopic Approach}

In contrast to photometry, spectroscopic analysis offers a more reliable way to determine multiple stellar parameters from just one observation. To evaluate the fundamental parameters and chemical composition of the programme stars, we employed a spectrum-fitting technique.

Our analysis utilised an \textsc{IDL}-based implementation of the \textsc{Spectroscopy Made Easy} package \citep[SME;][]{1996A&AS..118..595V, 2017A&A...597A..16P} with a grid of built-in \textsc{Atlas9} models to fit synthetic spectra to medium- and high-resolution observational data. Line lists were obtained from the Vienna Atomic Line Database \citep[VALD;][]{1995A&AS..112..525P, 2000BaltA...9..590K, 2015PhyS...90e4005R} based on \teff\ and \logg\ values estimated from photometry.

Taking the photometric \teff, \logg, and \mh\ as starting parameters, we refined \teff\ and \mh\ by fitting the LAMOST (HD\,23734 and HD\,73345) and XSHOOTER (HD\,68703) spectra. We preferred this approach to the direct handling of high-resolution data because the Balmer lines in \'echelle spectra of early-type stars typically span a significant fraction of a single order, complicating accurate continuum normalisation.

\begin{figure}
\centering
\includegraphics[width=\columnwidth]{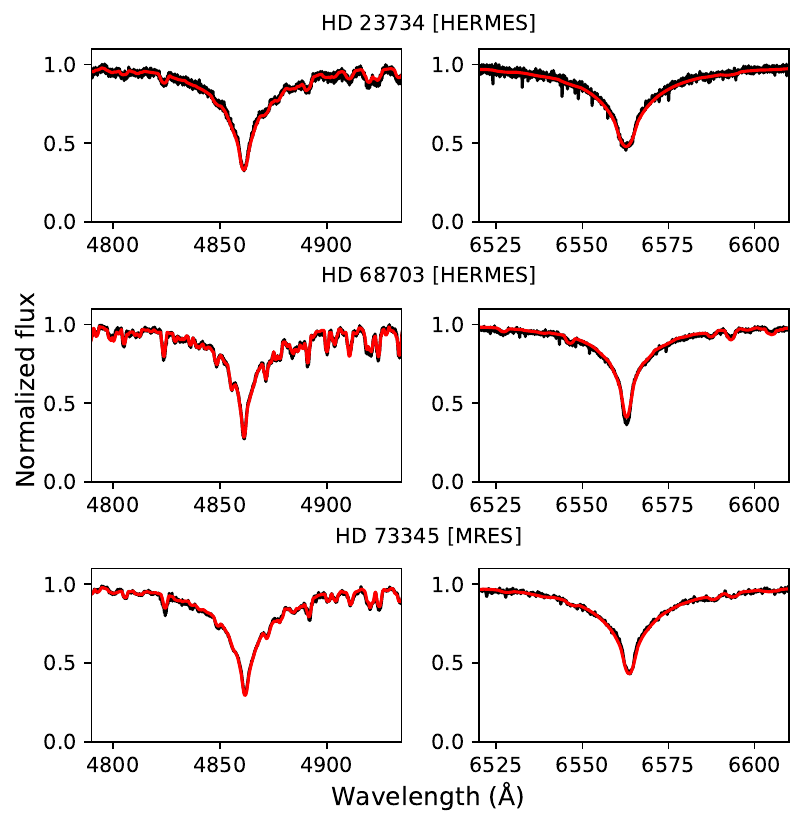}
\caption{ The H$_\beta$ (left panel) and H$_\alpha$ (right panel) line regions for the target stars. The spectrographs used to obtained these profiles are listed in the top of each panel.  The observed and synthetic profiles are shown with black and red colors, respectively.}
\label{fig:best_fit_spec}
\end{figure}

In the subsequent step, we analysed the high-resolution spectra to determine the surface gravity (\logg), microturbulence ($\xi_{\rm mic}$), and individual elemental abundances. To do so, \teff\ was fixed, \mh\ was set to zero, and hydrogen lines were excluded from the fit. All steps of analysis were performed in the Local Thermodynamic Equilibrium (LTE) approach. The final best-fit spectroscopically-derived parameters are presented in Table~\ref{tab:spec_params}.  To demonstrate the goodness-of-fit, for all the target stars, by the black solid lines in Fig. \ref{fig:best_fit_spec}, we plot the regions of spectra obtained with HERMES (HD\,23734, HD\,68703), and MRES (HD\,73345) and centred at the Balmer lines H$_\alpha$ and H$_\beta$. Spectra synthesised with the spectroscopically derived parameters are shown there as red lines. Within the error limits, the spectroscopic parameters agree with the photometric values listed in Table \ref{tab:ubvbeta}.

\begin{table}
\centering 
\caption{Summary of stellar parameters derived from spectroscopy.}
\label{tab:spec_params}
\fontsize{8}{10.0}\selectfont
\begin{tabular}{lccccc}
\hline
\hline
Star      & $ T_{\rm eff}$ & $\log g$        & $\xi_{\rm mic}$   & \mh         \\
          & (K)            & [cm\,s$^{-2}$]     & (\kms)         &  (dex)      \\
\hline
HD\,23734 &   $7670\pm160$  &  $3.98\pm0.15$ &   $2.7\pm0.5$ &  $-0.31\pm0.10$ \\
HD\,68703 &   $7160\pm250$  &  $3.40\pm0.15$ &   $2.7\pm0.3$ &  $+0.12\pm0.10$  \\
HD\,73345 &   $7670\pm220$  &  $4.08\pm0.15$ &   $3.0\pm0.3$ &  $-0.14\pm0.12$ \\ 
\hline\hline
 \end{tabular}
\label{table2}
\end{table}

\subsubsection{Stellar Rotation}

The rotational broadening of spectra was estimated using the Least-Squares Deconvolution (LSD) method \citep{donati1997}. The line mask required for deconvolution was constructed from VALD line lists, tailored to the derived fundamental parameters of each star. LSD profiles were fitted with a function describing rotational broadening \citep{Gray_2005}, with some of the best-fit profiles shown in Fig.~\ref{fig:LSD}. Two parameters, \vsini\ and RV, found in this way are listed in columns 10 and 11 of Table~\ref{tab:speclog}. From \textsc{sme} spectrum fitting, the averaged \vsini\ values for HD\,23734, HD\,68703, and HD\,73345 are $134\pm5$\,\kms, $69\pm2$\,\kms, and $88\pm3$\,\kms, respectively.

The LSD profiles of HD\,23734 and HD\,73345 exhibit minimal line profile variability (LPV), likely due to high-order pulsation combined with significant rotational broadening. In contrast, HD\,68703 shows substantial LPV which is confidently attributable to pulsational activity, making this star an excellent candidate for detailed follow-up studies.

In photometric time-series, we searched for the rotational frequencies in the range 0.1 to 5\,\cd\ \citep{10.1111/j.1365-2966.2011.18813.x} as shown in Fig.~\ref{fig:v_vsini}. In this diagram, viable rotational frequencies correspond to positions below the diagonal line representing 90$^{\circ}$ inclination, as \vsini\ must always be less than or equal to the equatorial velocity. For HD\,23734, we identified a frequency of 2.3406\,\cd, which exhibits one harmonic. Although its S/N ratio 
falls below the detection threshold of SNR~= 5.2, we classify this as a probable rotational frequency. In the case of HD\,68703, a frequency of 1.0995\,\cd\ was detected with two harmonics. For HD\,73345, we observed a frequency of 1.1119\,\cd, although no harmonics were identified. 
To evaluate the feasibility of the identified rotational frequencies, we estimated equatorial rotational velocities from the computed stellar radii and compared them to the spectroscopic \vsini\ values, indicating that these rotational frequencies are physically plausible. The inferred inclinations are approximately 45$^{\circ}$ for HD\,23734 and HD\,73345, while HD\,68703 likely has a lower inclination of $\sim$25$^{\circ}$, which requires additional verification.

\subsubsection{Atmospheric Abundances}

\begin{figure*}
\centering
\includegraphics[width=0.8\linewidth]{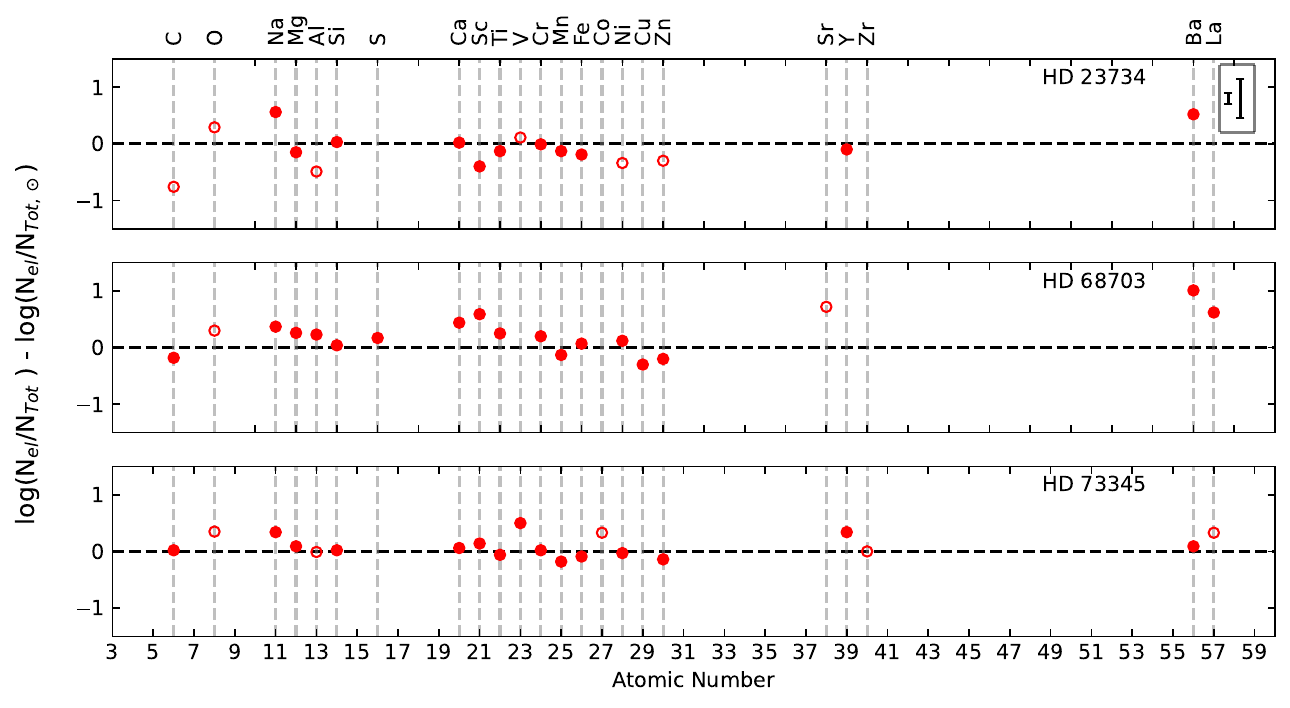}
\caption{Relative abundances of elements derived from spectroscopic analysis. The horizontal black dashed line indicates solar abundances, while the vertical grey dashed lines mark the one-to-one correspondence between names of the element and respective atomic numbers. The uncertain abundances are designated by the open circles. The box in the upper right corner shows the range (min-max) of the typical error bars ($\sim$ 0.10 to 0.35 dex).}
\label{abundance:fig}
\end{figure*}

The photospheric abundances of all dominant elements were evaluated from fitting multiple segments of spectra taken with HERMES (HD\,23734, HD\,68703) and MRES (HD\,73345) and summarised in Table~\ref{table_abundances}. In Fig.~\ref{abundance:fig} we plotted the chemical composition of stars relative to the Sun with the reference values adopted from \citet{2021A&A...653A.141A} and \citet{2022A&A...661A.140M}. We characterise the overall composition of HD\,23734 and HD\,73345 as being mildly metal-deficient, unlike the third star, HD\,68703, which shows a slight excess of many elements. The revealed difference in metallicity may reflect the individual specific properties of each star.

 In the present study, we ignored the effect of departures from the LTE. This fact might stand behind the abnormal excess of oxygen, which was measured from the O\textsc{i} 7771-7775\,\AA\ triplet \citep[e.g.][]{2013AstL...39..126S}. Even though the goodness-of-fit is high for the oxygen lines, we marked the derived abundance of this element as uncertain. The low accuracy of the rest of the values marked with a colon in Table \ref{table_abundances} mostly resulted from the weakness of measured lines, their limited number, or blending with more abundant species in rotationally broadened profiles.

\begin{table}
\centering
\caption{Individual chemical abundances determined for the target stars using HERMES (HD\,23734 and HD\,68703), and MRES (HD\,73345) spectra. The last column lists the solar abundances \citep{2021A&A...653A.141A, 2022A&A...661A.140M}. The abundances are expressed as $\log (N_{\rm el}/N_{\rm tot})$ with typical uncertainties ($\sigma(\log N/N_{\rm tot})$) between 0.10 and 0.35\,dex, except for least accurate values marked with a colon (:).}
\begin{tabular}{lcccccccc}
\hline\hline
     & HD\,23734    & HD\,68703   & HD\,73345   & Sun     \\
\hline
C   &  $-4.23$ (:)  & $-3.65$     & $-3.45$     & $-3.47$  \\
O   &  $-2.97$ (:)  & $-2.96$ (:) & $-2.91$ (:) & $-3.26$  \\
Na  &  $-5.25$      & $-5.44$     & $-5.47$     & $-5.81$  \\
Mg  &  $-4.63$      & $-4.22$     & $-4.39$     & $-4.48$  \\
Al  &  $-6.09$ (:)  & $-5.37$     & $-5.61$ (:) & $-5.60$  \\
Si  &  $-4.41$      & $-4.40$     & $-4.42$     & $-4.44$  \\
S   &  -            & $-4.74$     & -           & $-4.91$  \\
Ca  &  $-5.64$      & $-5.22$     & $-5.60$     & $-5.66$  \\
Sc  &  $-9.29$      & $-8.30$     & $-8.75$     & $-8.89$  \\
Ti  &  $-7.19$      & $-6.81$     & $-7.12$     & $-7.06$  \\
V   &  $-8.02$ (:)  & -           & $-7.63$     & $-8.13$  \\
Cr  &  $-6.42$      & $-6.21$     & $-6.39$     & $-6.41$  \\
Mn  &  $-6.74$      & $-6.74$     & $-6.79$     & $-6.61$  \\
Fe  &  $-4.71$      & $-4.45$     & $-4.61$     & $-4.52$  \\
Co  &  -            & -           & $-6.76$ (:) & $-7.09$  \\
Ni  &  $-6.13$ (:)  & $-5.67$     & $-5.82$     & $-5.79$  \\
Cu  &  -            & $-8.15$     & -           & $-7.85$  \\
Zn  &  $-7.77$ (:)  & $-7.67$     & $-7.61$     & $-7.47$  \\
Sr  &  $-11.05$ (:) & $-8.48$ (:) & -           & $-9.20$  \\
Y   &  $-9.92$      & $-9.26$     & $-9.48$     & $-9.82$  \\
Zr  &  -            & -           & $-9.44$ (:) & $-9.44$  \\
Ba  & $-9.24$       & $-8.75$     & $-9.67$     & $-9.76$  \\
La  & -             & $-10.30$    & $-10.59$ (:) & $-10.92$ \\
\hline\hline
\end{tabular}
\label{table_abundances}
\end{table}

 Our analysis indicates that neither of the programme stars shows strong anomalies typical for CP stars. This conclusion complements the negative test for the presence of a flux depression around 5200\,\AA~\citep{2004MNRASKupka} in spectra taken with low resolution. We checked for such features in the flux-calibrated low-resolution spectra obtained with LAMOST for HD\,23734 and HD\,73345, and a medium-resolution spectrum of HD\,68703 taken with XSHOOTER. Unfortunately, a flux-calibrated LAMOST spectrum of HD\,68703 is unavailable in the LAMOST archive. For comparison, we used {\it bona fide} CP1 and CP2 stars with a spectral type comparable to that of the target stars. On inspecting the spectra shown in Fig. \ref{fig:flux_dep}, one can notice that HD\,23734 and HD\,73345 do not possess any anomalies near 5200\,\AA. Furthermore, CP1 stars usually exhibit a distinctly weak Ca II K line in their spectra. This line in the spectra of our targets appears normal (Fig.~\ref{fig:ca2k}), thereby confirming that neither of the studied stars belongs to the Am (CP1) class. In summary, our study portrays HD\,23734, HD\,68703, and HD\,73345 as chemically normal stars.

\subsubsection{Stellar Multiplicity} \label{multiplicity}

We measured the radial velocities (RV) from the available spectra as listed in Table~\ref{tab:speclog}. Among the three stars, notably, only HD\,73345 showed significant variation of RV over time, suggesting it may be part of a long-period binary system. However, due to the limited temporal sampling of the available data, we were unable to determine the orbital period with confidence, and additional spectroscopic follow-up observations might therefore be essential to confirm the binary status of HD\,73345.

In contrast, HD\,68703 displays RV variations with a standard deviation of approximately 0.7\,\kms, consistently with rather pulsational variability than binarity. We therefore classify this object as a single star.


\subsection{Evolutionary Status} \label{evstat}

To know the evolutionary status of the sample stars, we placed them in the H–R diagram. For this, we used luminosities derived from \gaia\ parallaxes and effective temperatures (\teff) obtained from high-resolution spectroscopy. The absolute magnitudes ($M_{\rm V}$) were calculated from the \gaia\ parallaxes \citep{2016A&A...595A...2G} using the standard relation \citep{2000asqu.book.....C}. To compute the stellar luminosity (\logL), we adopted a solar bolometric magnitude of $M_{\rm bol,\odot} = 4.73$\,mag \citep{2010torres}. The bolometric correction (BC) was determined using an empirical relation taken from \citet{2010torres}. The resulting values of $M_{\rm V}$ and \logL, along with their associated uncertainties, are listed in Table\,\ref{tab:ubvbeta} and \ref{tab:par_tab}, respectively.

\begin{figure}
\centering
\includegraphics[width=\columnwidth]{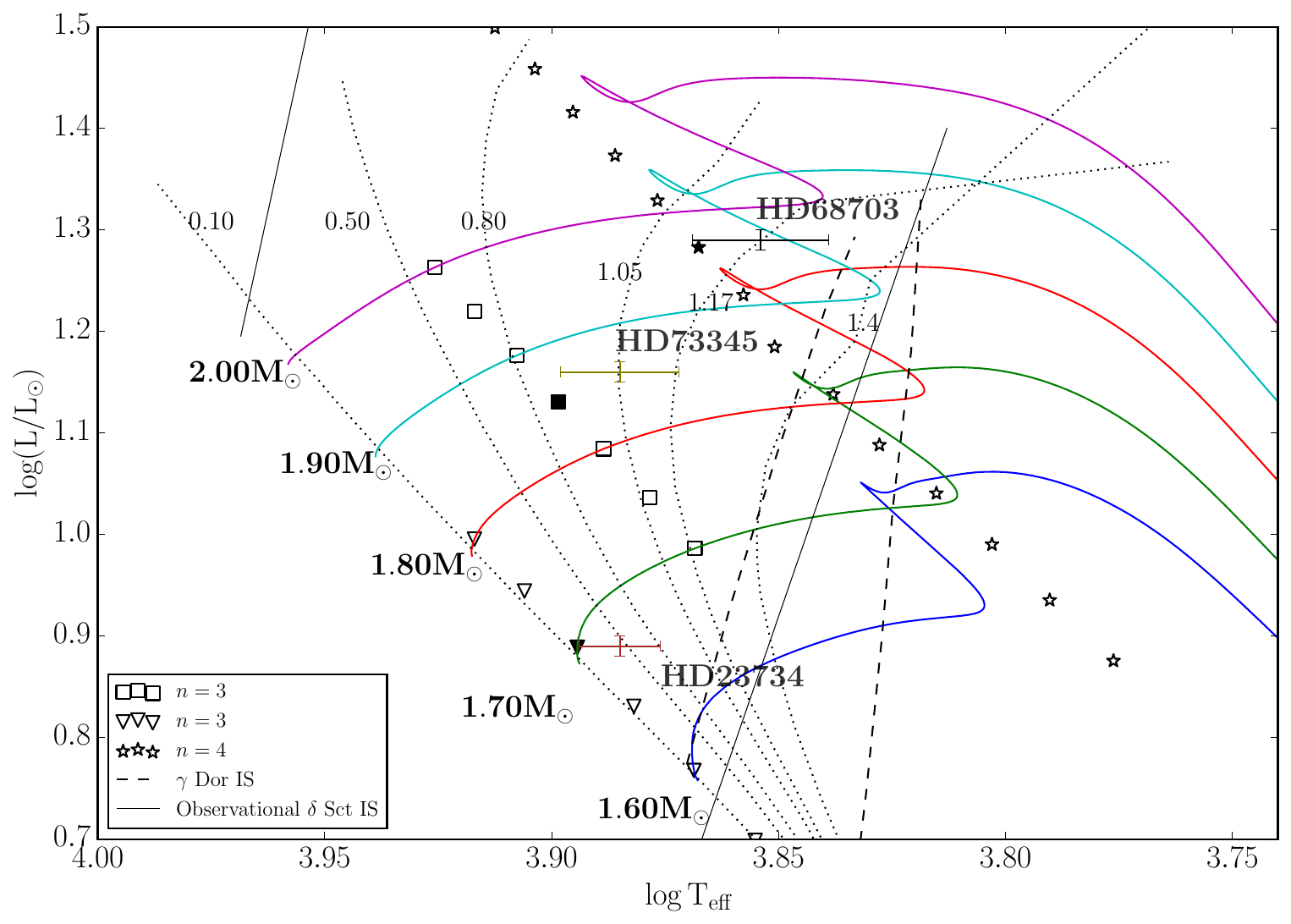}
\caption{
Evolutionary tracks for stellar masses ranging from 1.60\,\mdot\ to 2.00\,\mdot, computed for solar metallicity ($Z=0.014$) and overshooting parameter $\alpha_{\rm ov}=0.1$. Dotted lines denote isochrones corresponding to ages of 0.1, 0.5, 0.8, 1.05, 1.17 and 1.4\,Gyr. The open symbols represents seismic models minimizing $\chi^2$ for radial overtone modes: $n=3$, $f_{T7}=32.94\,\mathrm{d}^{-1}$ for HD\,23734 (triangle); $n=4$, $f_{T1}=18.58\,\mathrm{d}^{-1}$ for HD\,68703 (star); and $n=3$, $f_{T12}=24.72\,\mathrm{d}^{-1}$ for HD\,73345 (square) where each model is computed for different stellar masses. The models are consistent with the observations within the 1$\sigma$ uncertainty box, except for HD\,73345 which is likely to be part of a binary system. 
The solid black lines represent the observational \dsct\ instability strip, while the dashed black lines denote the theoretical instability strip of the $\gamma$\,Doradus stars.
}
\label{HR}
\end{figure}

The locations of HD\,23734, HD\,68703, and HD\,73345 in the H–R diagram are shown in Fig.\,\ref{HR}. It is found that position of HD\,23734 is near to the zero-age main sequence (ZAMS) indicates it has recently begun core hydrogen burning. HD\,68703 lies near the terminal-age main sequence (TAMS), suggesting that hydrogen in the core is nearly exhausted. HD\,73345 is positioned between the ZAMS and TAMS and is likely still undergoing  core hydrogen burning. 

We have overplotted the stellar evolutionary tracks for masses ranging between 1.6 and 2.0\,\mdot, and ages ranging from 0.1 to 1.4\,Gyr, computed using the stellar evolution code \textsc{CL\'{E}S} (Code Li\'{e}geois d’Evolution Stellaire; \citealt{2008Ap&SS.316...83S}). Theoretical \gdor\ \citep{2004A&A...414L..17D} and empirical \dsct\ \citep{murphyis2019} instability strips are overlaid for reference. An inspection of the stars' locations within the instability strips, in combination with their frequency spectra, confirms that HD\,23734, HD\,68703, and HD\,73345 are \dsct-type pulsating variables. It is important to note that the observed values of the luminosity of HD\,73345 can be slightly higher than the actual luminosity of the star due to its possible membership to a binary system. If the star has a companion, the combined flux could cause an observed vertical displacement in the H–R diagram, placing HD\,73345 above its true evolutionary position.

\section{Mode Identification} \label{modes}

Asteroseismology leverages stellar pulsation frequencies and their corresponding amplitudes to infer fundamental stellar properties, thereby contributing significantly to our understanding of stellar structure and evolution. A crucial prerequisite for applying asteroseismic techniques is the accurate identification of pulsation modes, which remains a challenging task even when high-precision space-based data are available.
\citep{2010aste.book.....A}. In the following subsections, we outline several approaches employed for mode identification in our target stars.

\begin{figure*}
\centering
\includegraphics[width=0.33\textwidth]{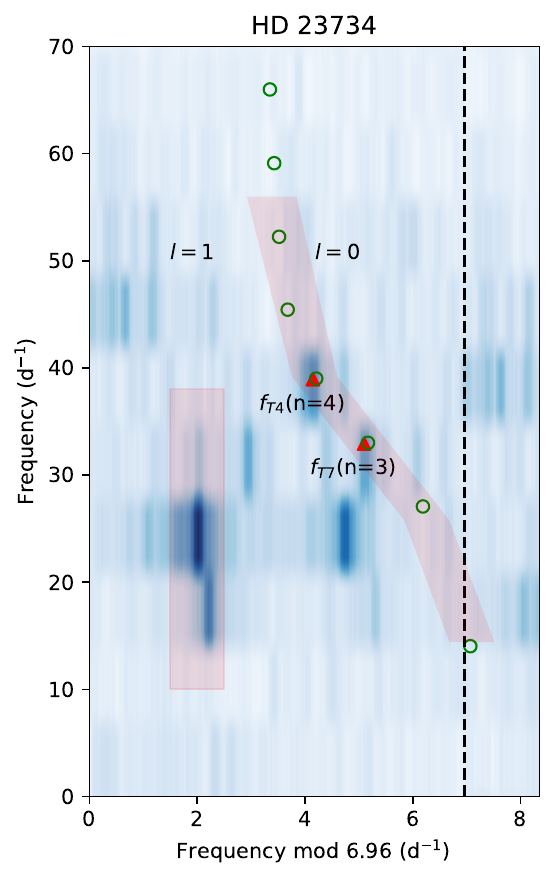}
\includegraphics[width=0.33\textwidth]{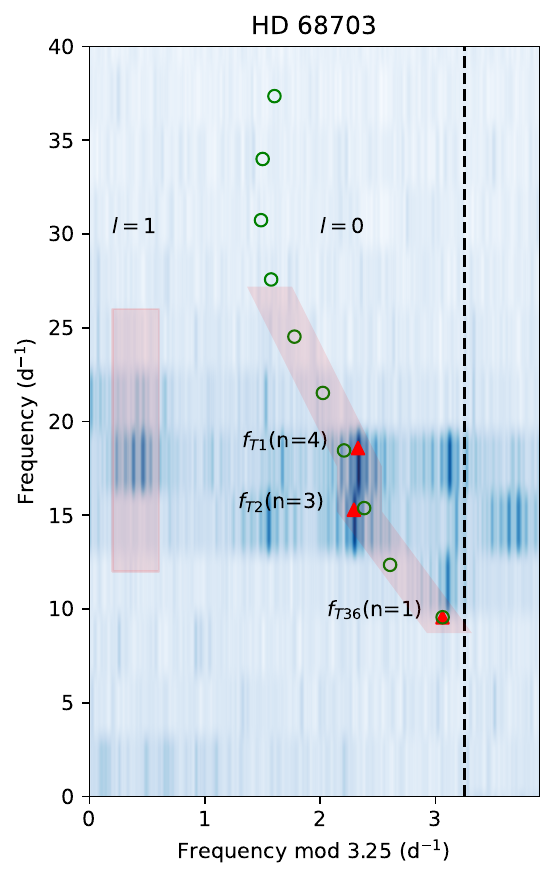}
\includegraphics[width=0.33\textwidth]{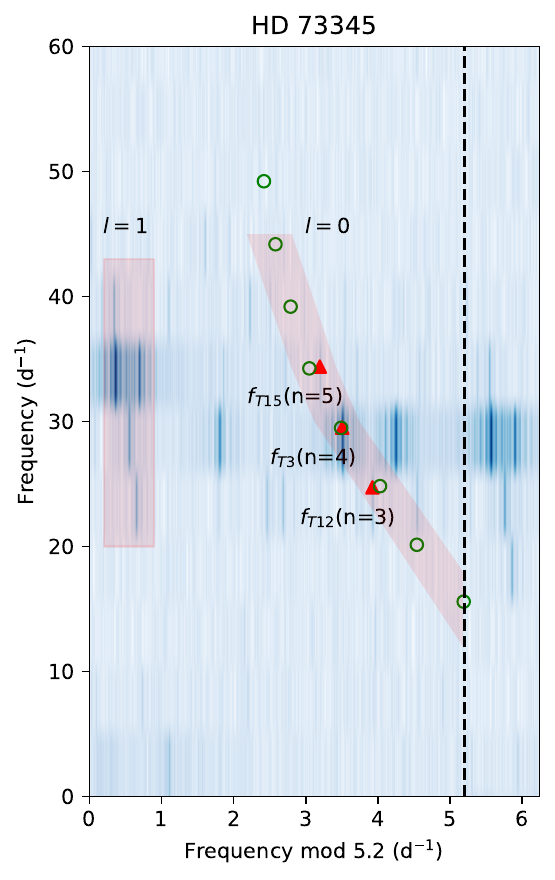}
\caption{
\'{E}chelle diagrams of HD\,23734 (\textit{left panel}), HD\,68703 (\textit{middle panel}), and HD\,73345 (\textit{right panel}), constructed using their respective large frequency separations of $\Delta \nu$ = 6.96, 3.25, and 5.20\,\cd\ (indicated by vertical lines). For clarity, the frequency range is extended by 20\% beyond the nominal value of $\Delta \nu$. Radial modes predicted by seismic models are shown as open green circles, while observed frequencies are plotted as filled red triangles. The \tess\ frequencies and corresponding radial overtones are denoted by $f_T$ and $n$, respectively. A notable feature of the \'{e}chelle spectra for low- and intermediate-mass pulsators is the appearance of vertical ridges \citep{2020Tim}, which are visible in all three targets and highlighted by the light red shaded regions.}
\label{fig:ech-hd68703}
\end{figure*}

\subsection{\'{E}chelle Diagram}

One of the robust techniques for identifying pulsation modes in asteroseismology is the \'{E}chelle diagram (ED). In an ED, for a given large frequency separation ($\Delta\nu$), the observed frequencies corresponding to the same angular degree ($\ell$) are folded into segments. Frequencies associated with low angular degrees ($\ell < 2$) and high radial orders ($n \gg \ell$) of pressure modes approximately follow the asymptotic relation \citep{1980ApJS...43..469T}:
\begin{equation}
    \nu_{n,\ell} = \Delta\nu (n + \ell/2 + \epsilon),
    \label{eq:asymptotic}
\end{equation}
where $n$ is the radial overtone number, $\ell$ is the angular degree, and $\epsilon$ is a constant of order unity. For higher radial orders, modes of the same degree and consecutive overtone numbers are nearly equally spaced in frequency \citep{2020Tim}, and this regularity manifests as vertical ridges in the ED.

The large frequency separation $\Delta\nu$ is empirically related to the star's mean density ($\bar\rho$) via the following relation \citep{2014A&A...563A...7S}:
\begin{equation}
    \frac{\Delta\nu}{\Delta\nu_{\odot}} = 0.776 \left(\frac{\bar\rho}{\bar\rho_{\odot}}\right)^{0.46},
\end{equation}
where $\Delta\nu_{\odot} = 135\,\mu$Hz $\approx 11.67$\,\cd\ is the solar large frequency separation \citep{2011ApJ...743..143H} and $\bar\rho_{\odot}$ is the mean density of the Sun. To obtain an initial estimate of $\Delta\nu$, we computed the mean stellar density using $\bar\rho = M / R^3$, where the radius $R$ was derived from the Stefan–Boltzmann law ($L \propto R^2 T_{\rm eff}^4$) and the mass $M$ was estimated from theoretical evolutionary tracks. These values were then used to refine $\Delta\nu$ through the dynamic \'{e}chelle interface \citep{2022hey}.

Fig.~\ref{fig:ech-hd68703} shows the \'{E}chelle diagrams of HD\,23734, HD\,68703, and HD\,73345, with $\Delta \nu$ = 6.96, 3.25, and 5.20\,\cd, respectively. We identified radial ($\ell = 0$) and possibly non-radial ($\ell = 1$) ridges for all three samples and are highlighted in the figure with light red shaded regions.

\subsection{Seismic Modelling}\label{seismodel}
Seismic modelling is an independent method for accurately measuring the age of stars based on observed frequencies. The conventional approach of employing theoretical isochrones presents considerable uncertainty, particularly near the Zero-Age Main Sequence (ZAMS), where the theoretical isochrones are closely spaced.

We adopted an iterative approach to identify the radial pulsation modes using theoretical models. Our working hypothesis assumed that the highest amplitude frequencies ($\nu^{\rm radial}_{\rm max}$) correspond to radial modes of an arbitrary radial order ($n$) in evolved main sequence stars, while they correspond to non-radial modes in less evolved stars with a higher surface gravity. To investigate this, we employed the stellar evolution code \textsc{CL\'{E}S} (Code Li\'{e}geois d’Evolution Stellaire; \citealt{2008Ap&SS.316...83S}) to compute evolutionary models with stellar masses ranging from $1.40$ to $2.20\,M_{\odot}$, assuming solar metallicity ($Z = 0.014$) and a convective core overshooting parameter of $\alpha_{\rm ov} = 0.1$.
For each model, we computed the radial mode frequencies ($\ell = 0$) using the adiabatic stellar oscillation code \textsc{OSC} \citep{2008Ap&SS.316..149S}, considering various values of the radial overtone number $n$ associated with $\nu^{\rm radial}_{\rm max}$. The most likely $n$ value was determined by minimizing the seismic-$\chi^2$, following the formulation of \citet{2021MNRAS.502.1633M} :
\begin{equation}
\chi^{2} = \left( \frac{f_{t} - f_{o}}{\sigma_{f_o}} \right)^2,
\end{equation}
where $f_{t}$ and $f_{o}$ are the theoretical and observed frequencies of the identified radial mode, respectively, and $\sigma_{f_o}$ is the uncertainty in the observed frequency, taken as 0.1\,d$^{-1}$ \citep{2021MNRAS.502.1633M}.
The best-fitting models for each star were selected based on the minimum $\chi^2$ value, and are overplotted on the H-R diagram and is shown in Fig.~\ref{HR}, alongside evolutionary tracks for various stellar masses. The derived seismic parameters for HD\,23734, HD\,68703, and HD\,73345 are listed in Table~\ref{tab:par_tab} and tagged with column \textit{Seismology}.

The identified radial modes that align closely with the models are the second and third overtones for HD\,23734; the fundamental, second, and third overtones for HD\,68703; and the second, third, and fourth overtones for HD\,73345.

\begin{table*}
\centering
\scriptsize
\caption{Estimated and calculated fundamental parameters derived from the evolutionary status, inferred from the \'{E}chelle diagram and best-fitting asteroseismic models for HD\,23734, HD\,68703, and HD\,73345. 
The parameters reported in the literature are also provided for comparison to the parameters inferred from the present study. 
The identified frequencies  $Fn$  corresponds to the radial fundamental ($n = 1$) and higher overtones ($n > 1$) modes.}
\label{tab:par_tab}
\begin{tabular}{|cccccccccc|}
\hline\hline
            &            & HD\,23734   &         &            & HD\,68703   &         &            & HD\,73345   &         \\
\hline
& \multicolumn{2}{c|}{Present Study} &  & \multicolumn{2}{c|}{Present Study} &  & \multicolumn{2}{c|}{Present Study} &  \\
\cline{2-3} \cline{5-6} \cline{8-9}
& \multicolumn{2}{c|}{} &  & \multicolumn{2}{c|}{} & & \multicolumn{2}{c|}{} & \\

          Parameters  & Calculated/   & Seismology & Literature  & Calculated/  &  Seismology &Literature  & Calculated/   & Seismology &  Literature \\
  & H-R diagram &  &   & H-R diagram      &  &    & H-R diagram      &  &   \\
\hline
$\Delta\nu$ (\cd)             & 6.17 $\pm$0.47 &   6.96 &    & $3.00\pm0.33$ &  $3.25$ &    & $4.14\pm0.41$ & $5.20$   & 5.01$^b$  \\
 
$\bar{\rho}/\bar{\rho}_{\odot}$ & 0.43$\pm$0.07 &0.57  &    & $0.09\pm0.02$ &   $0.11$  &    & $0.18\pm0.04$ &  $0.30$  &    \\
$R_{L}/R_{\odot}$               & 1.57$ \pm$0.08 & -    & -     & $2.79\pm0.22$ & -    & -     & $2.17\pm0.15$ & -    & -     \\
$M/M_{\odot}$                  &1.67 $\pm$0.10 & -    & -     & $1.96\pm0.10$ & -    & -     & $1.85\pm0.10$ & -    & -     \\
\logL                        & $0.89\pm0.01$      & 0.841 $\pm$0.050 & -& $1.27\pm0.01$ &   $1.175\pm0.100$ & 1.29$^d$ & $1.17\pm0.01$ &      $1.120\pm0.050$ & 1.17$^g$ \\
$\log$(\teff)                & 3.88$\pm$0.01      & 3.881$\pm$0.010 & -& $3.85\pm0.02$ &    $3.854\pm0.020$ & 3.86$^d $&$3.88\pm0.01$ &   $3.899\pm0.010$ & 3.90$^g$\\
Age (Gyr)                    & 0.5 $\pm$ 0.3       & 0.327$\pm$0.010 & 0.832$^c$ & $1.17\pm0.10$   &   $1.290\pm0.200$ & - & $1.05\pm0.10$   &      $0.863\pm0.100$ & 0.759$^{a}$ \\  
$\rm [Fe/H]$             & $-0.19 $  &  - & -0.12$^c$ & $0.07 $ & - &0.352$^f$ &$-0.09 $&- & 0.26$^g$ \\

F1 (\cd)                     & -        & - & & 9.5603 $\pm$ 0.0007     &  $9.56\pm0.10$  & -             & -    &  -    \\
F2 (\cd)                     & -                  & -     & -             & -    & -     & -             & -    & -     & -\\
F3 (\cd)                     & 32.93878 $\pm$ 0.00009       & 33.01 $\pm$ 0.50& - & 15.29221 $\pm$ 0.00002   &    $15.38\pm0.20$ &- & 24.7246 $\pm$ 0.0005    &  $24.83\pm0.02$ & -\\
F4 (\cd)                     & 38.94782 $\pm$ 0.00006         & 39.01 $\pm$ 0.40 & - & 18.57798 $\pm$ 0.00001   &   $18.46\pm0.20$ & -& 29.50692 $\pm$ 0.00005   &   $29.49\pm0.01$ & -\\
F5 (\cd)                     & -         & - & -             & -    & - &    & 34.3951 $\pm$ 0.0007    &  $34.25\pm0.80$ & -\\
\hline
\hline
\end{tabular}
\begin{minipage}{\textwidth}      
    \small                          
     $^{a}$\citet{2023A&A...677A.163A},
    $^{b}$\citet{2023A&APamos},
    $^c$\citet{2021MNRASDias},
    $^d$\citet{2024A&A...690A.104D},
    $^f$\citet{2020ApJS..251...15Z},
    $^g$\citet{2008A&A...483..891F}
\end{minipage}

 \end{table*}
\subsection{Frequency Ratios}

\begin{figure*}
\includegraphics[width=0.45\linewidth]{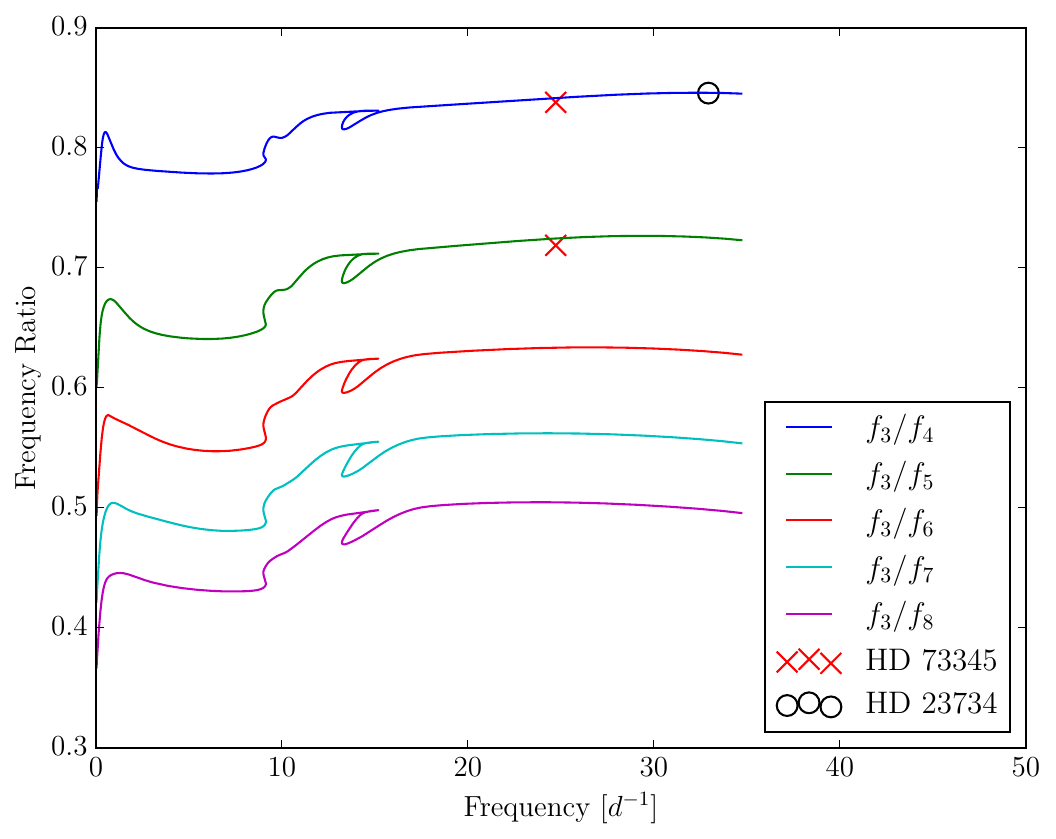}
\includegraphics[width=0.45\linewidth]{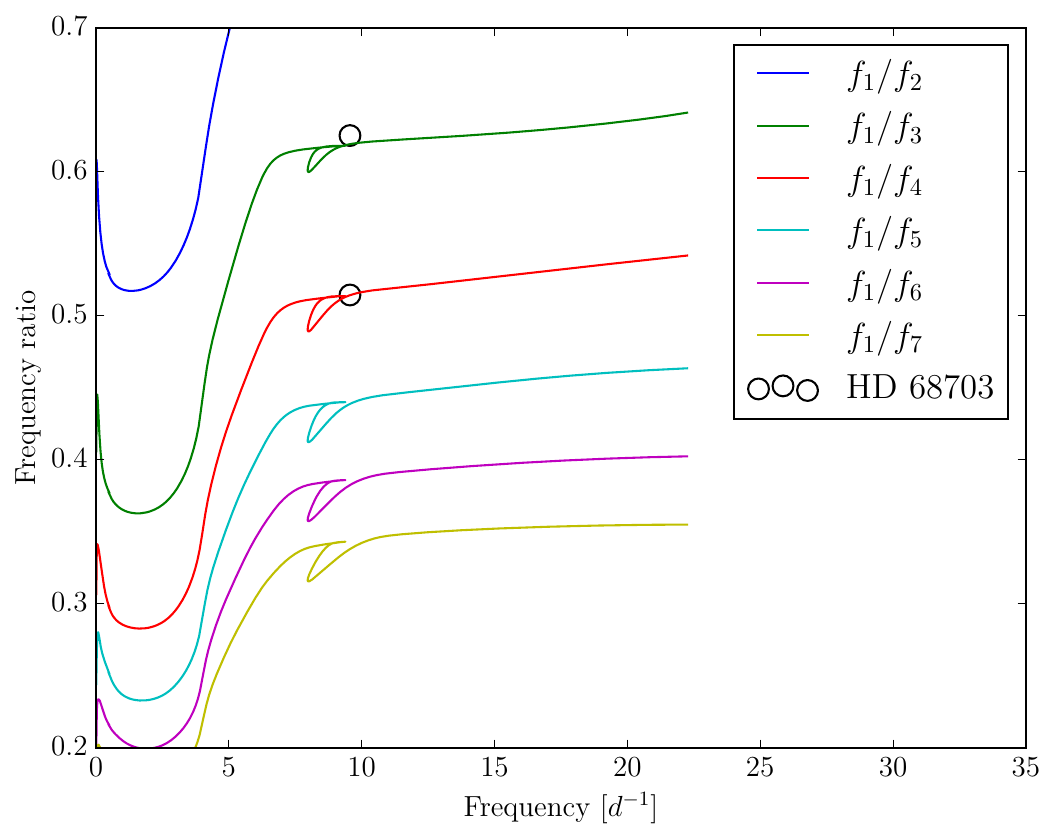}

\caption{
  Petersen diagram for HD\,23734 and HD\,73345 (\textit{left panel}) and HD\,68703 (\textit{right panel}), computed for $Z=0.014$ and $\alpha_{\rm ov}=0.1$. The crosses and circles represent the frequency ratios of the second overtone ($f_3$) to higher radial overtones for HD\,23734 and HD\,73345, respectively. For HD\,68703, the frequency ratios are calculated for the fundamental  ($f_1$) to higher order modes. The modes identified via frequency ratios show good agreement with those inferred from the \'{E}chelle diagrams.
}
\label{fig:petersen}
\end{figure*}

When multiple radial modes are detected in a pulsating star, the Petersen diagram \citep{1973A&A....27...89P, 1978A&A....62..205P}, that depicts the ratio of the higher to lower radial mode frequencies of a star, is generally used to verify the identifications. In this analysis, we compared the theoretical frequency ratios of the higher radial overtones ($n \geq 2$) to the lower radial modes or the fundamental radial mode ($n = 1$) if present. Fig.\,\ref{fig:petersen} presents the Petersen diagram for our target stars.
These results are in agreement with the mode identifications obtained through the \'{E}chelle diagrams and the seismic models.

\subsection{Pulsation Constant}

To validate the identified radial pulsation modes, we calculated the pulsation constant, $Q$ for each target star, following the formulation provided by \citet{1990DSSN....2...13B}. The $Q$ values for HD\,23734, HD\,68703, and HD\,73345 are listed in the fifth columns of Tables\,\ref{tab:tablet1}, \ref{tab:tablet2}, and \ref{tab:tablet3}, respectively. These values fall within the expected ranges for the fundamental and overtone modes, with fractional uncertainties up to 18\%. The adopted $Q$ ranges for radial modes are as follows: fundamental ($Q \geq 0.027$), first overtone ($0.021 \leq Q < 0.027$), second overtone ($0.018 \leq Q < 0.021$), and third overtone ($Q < 0.018$), as given by \citet{2024RAA....24b5011P}.
The sixth columns of these tables indicate the corresponding identified pulsation modes, labelled as fundamental ($F_1$) and successive overtones ($F_2$ to $F_5$). 
It is found that HD\,23734 pulsates in the second and third overtone, 
HD\,68703 shows pulsations in the fundamental, second, and third overtones and 
HD\,73345 exhibits radial pulsations in the second, third, and fourth overtones.
The agreement among the $Q$ values, \'{e}chelle diagrams, seismic models and frequency ratios supports the reliability of the identified radial pulsation modes.

\section{Conclusions and Future Prospects} \label{concl}

The primary goal of the extension of the N-C survey is to gain insights into the interplay between stellar pulsation, chemical inhomogeneities, rotation, and magnetic fields. To achieve the identified goals, we utilised state-of-the-art photometric and spectroscopic data acquired from both ground- and space-based facilities, followed by thorough analyses. Based on our investigation, we draw the following conclusions:

\begin{itemize}

\item[(i)] Analysis of \textit{K2} and \tess\ photometry reveals that HD\,68703 and HD\,73345 are multi-periodic \dsct\ pulsators, while HD\,23734 is a newly identified \dsct\ pulsator. In our studied stars, we have also observed evidence of the coexistence of rotational modulation in the presence of the pulsational variability.
\\

\item[(ii)] We have determined the basic astrophysical parameters of HD\,23734, HD\,68703, and HD\,73345 using photometric and spectroscopic data. The lack of anomalies in measured individual abundances, together with the absence of flux depression at 5200\,\AA\ in flux-calibrated spectra, confirms that the studied objects are likely to be chemically normal A- and F-type stars.
\\


\item[(iii)]

The stellar parameters of HD\,68703 derived by \citet{2024A&A...690A.104D} are generally consistent with our findings.
The Least Squares Deconvolution (LSD) profiles of HD\,68703 reveal notable line profile variations due to pulsation, making it an ideal candidate for follow-up time-series spectroscopic studies. \\

\item[(iv)]

One of the targets, HD\,73345, is a member of the Praesepe cluster \citep{2023A&APamos}, which has an estimated age of 759\,Myr. This value is relatively close to the 863\,Myr derived from our asteroseismic analysis. \citet{2023A&APamos} reported a large frequency separation of $\Delta\nu = 58 \mu$Hz (approximately 5.01 \cd) for HD\,73345, which is in good agreement with our dynamically derived value of $\Delta\nu = 5.20$ \cd. Further, our RV measurements suggests that HD\,73345 is a likely member of a binary system, albeit,  the limited phase coverage restricts to make a conclusive results, and demands additional spectroscopic observations to confirm and characterise the system.
\\

\item[(v)] 
Through our comprehensive asteroseismic analysis, we identified the observed radial modes for all three stars. HD\,23734 and HD\,73345 exhibit predominant non-radial modes, whereas HD\,68703 displays prominent radial modes. The observed radial pulsation modes were found to be of radial orders $n$ = 3 and 4 for HD\,23734; $n = 1, 3,$ and $4$ for HD\,68703; and $n = 3, 4,$ and $5$ for HD\,73345. In addition to the radial modes, we also detected the probable non-radial ridge of $\ell$=1.
 \\

\end{itemize}

Among the samples observed under the N-C Survey project using ground-based facilities, we have identified several targets that have also been observed by the \tess\ mission \citep{2024BSRSL..93..227D}. These stars were previously classified as non-pulsators based on ground-based observations in the N-C Survey. Therefore, we propose a re-examination of all the objects using high-precision space-based photometry and high-resolution spectroscopy. Thus, the present study is an important step towards finding clues to understand the interactions between pulsation, rotation, binarity, and chemical peculiarities in the intermediate-mass stars.

\section*{Data availability statement}

The high-resolution spectroscopic data used in this article will be shared upon reasonable request to the corresponding author. The \textit{TESS} time-series flux data for the target stars are publicly available from the NASA MAST archive (\url{https://mast.stsci.edu/portal/Mashup/Clients/Mast/Portal.html}). 
This work has made use of data from the European Space Agency (ESA) mission \textit{Gaia} (\url{https://www.cosmos.esa.int/gaia}), processed by the Gaia Data Processing and Analysis Consortium (DPAC, \url{https://www.cosmos.esa.int/web/gaia/dpac/consortium}). Funding for DPAC has been provided by national institutions, in particular those participating in the \textit{Gaia} Multilateral Agreement. 
This research has also made use of the SIMBAD database, operated at CDS, Strasbourg, France.

\bibliographystyle{mnras}
\bibliography{ref}
 
\section*{Acknowledgments}
 The authors acknowledge the anonymous referee for providing the insightful comments. 
This work is supported by the Belgo-Indian Network for Astronomy and Astrophysics (BINA), approved by the International Division, Department of Science and Technology (DST, Govt. of India; DST/INT/BELG/P-09/2017), and the Belgian Federal Science Policy Office (BELSPO, Govt. of Belgium; BL/33/IN12) (BINA).
SJ, SCG, and OT acknowledge the financial support received from the BRICS grant DST/ICD/BRICS/Call-5/SAPTARISI/2023(G).
AD acknowledges the financial support received from the DST-INSPIRE Fellowship Programme (DST/INSPIREFellowship/2020/IF200245).
OT acknowledges financial support from the Uppsala University International Science Programme (ISP), the SEISMIC project, and the Max Planck - Humboldt Research Unit, in collaboration with MPA, MPS, and Kyambogo University.
SJ, APY, and SP gratefully acknowledge the financial support from the Core Research Grant (CRG/2021/007772) of the Science and Engineering Research Board (SERB), India.
This paper includes data collected by the \kepler\ and \tess\ missions, available from the Mikulski Archive for Space Telescopes (MAST), operated by NASA.
Based on observations made with the Mercator Telescope, operated on the island of La Palma by the Flemish Community at the Spanish Observatorio del Roque de los Muchachos of the Instituto de Astrofísica de Canarias. Based on observations obtained with the HERMES spectrograph, which is supported by the Research Foundation – Flanders (FWO), Belgium; the Research Council of KU Leuven, Belgium; the Fonds National de la Recherche Scientifique (F.R.S.-FNRS), Belgium; the Royal Observatory of Belgium; the Observatoire de Genève, Switzerland; and the Thüringer Landessternwarte Tautenburg, Germany.
Guoshoujing Telescope (the Large Sky Area Multi-Object Fiber Spectroscopic Telescope, LAMOST) is a National Major Scientific Project built by the Chinese Academy of Sciences. Funding for the project has been provided by the National Development and Reform Commission. LAMOST is operated and managed by the National Astronomical Observatories, Chinese Academy of Sciences. The LAMOST spectra are available at \url{https://www.lamost.org/}.
Part of the research is based on data obtained from the European Southern Observatory (ESO) Science Archive Facility with DOI(s): \url{https://doi.org/10.18727/archive/24}, \url{https://doi.org/10.18727/archive/50}, and \url{https://doi.org/10.18727/archive/71} under ESO Prog ID 110.248M.001.
This research has used data, tools, or materials developed as part of the EXPLORE project, which has received funding from the European Union’s Horizon 2020 research and innovation programme under grant agreement No.~101004214.

\onecolumn
\appendix

\pagebreak

\section{TESS and K2 Light Curves and Amplitude Spectra}

\begin{figure*}
\centering
\includegraphics[width=\textwidth]{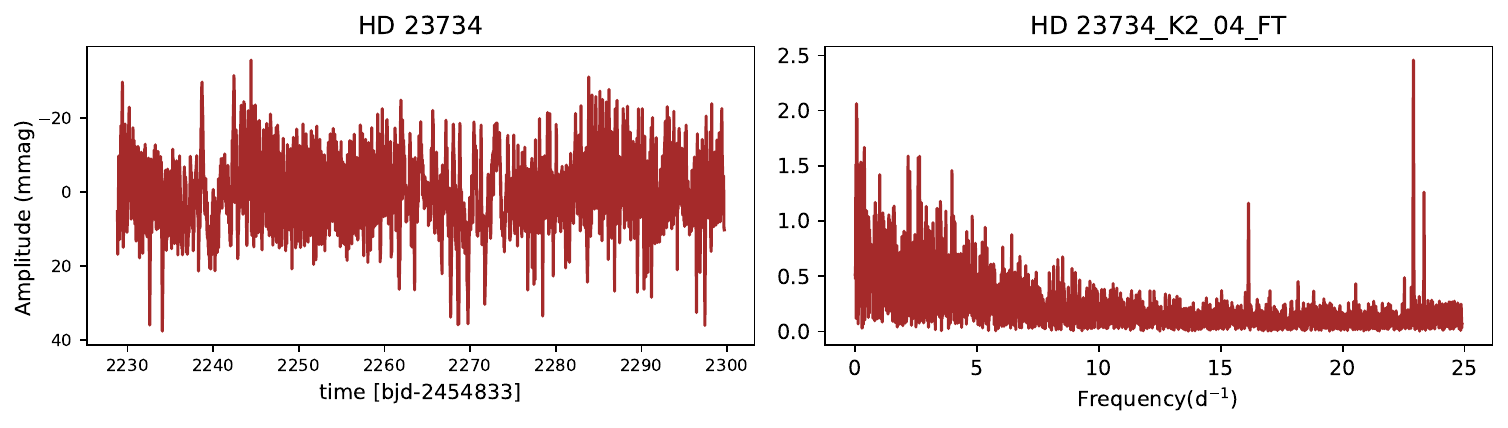}
\caption{The light curve (\textit{left panel}) and corresponding frequency spectrum (\textit{right panel}) for HD\,23734 obtained with \ktwo\ during Campaign 4.}
\label{fresp_k2_23734}
\end{figure*}

\begin{figure*}
\centering
 \includegraphics[width=\textwidth]{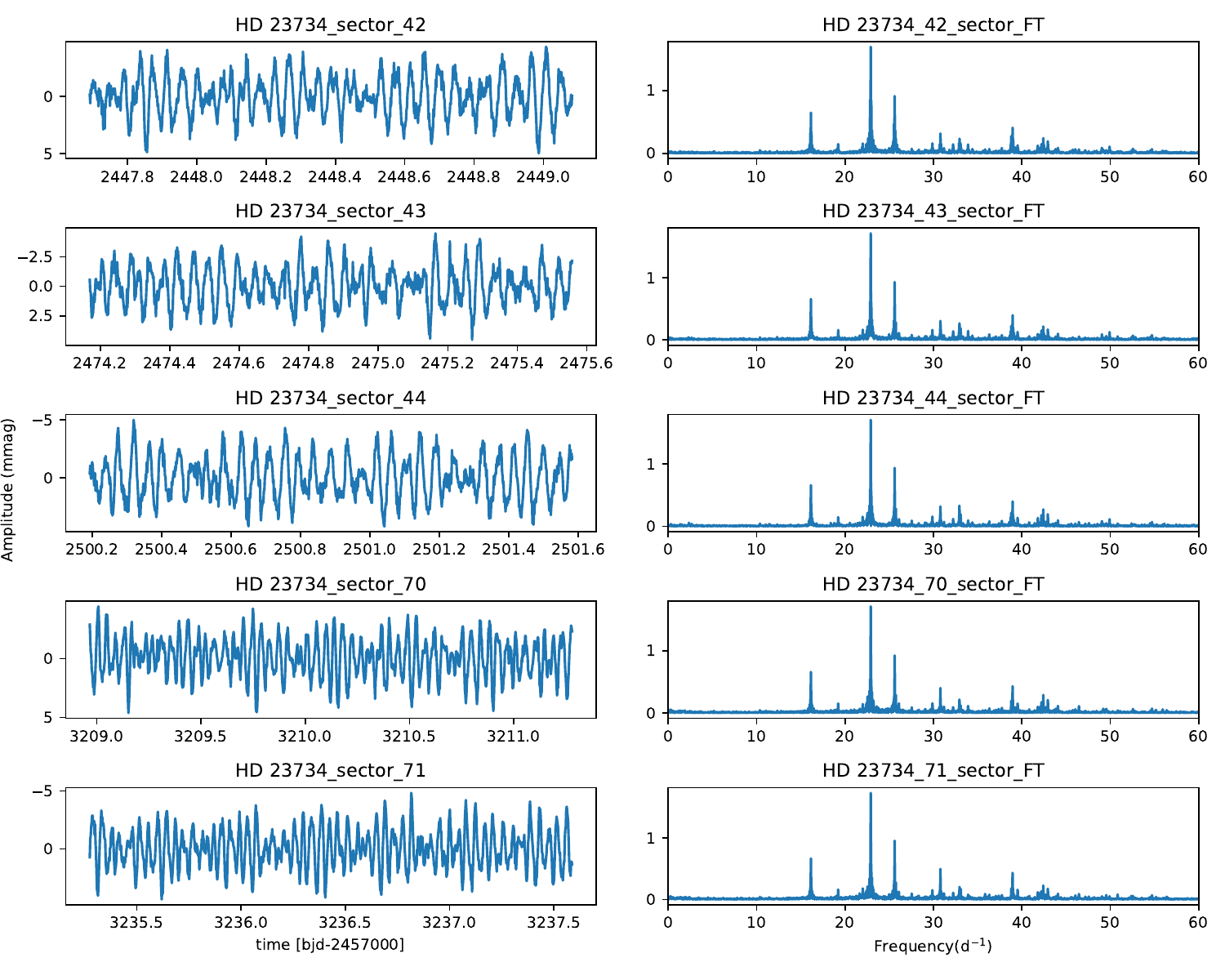}
\caption{
The light curves (\textit{left panels}) and corresponding frequency spectra (\textit{right panels}) for HD\,23734 obtained with \tess\ in Sectors 42, 43, 70, and 71 (\textit{from top to bottom row}). The frequency spectra are depicted only up to 60\,\cd\ as no significant frequencies were found beyond this point.
}
\label{fresp_tess_23734}
\end{figure*}

\begin{figure*}
\centering
\includegraphics[width=\textwidth]{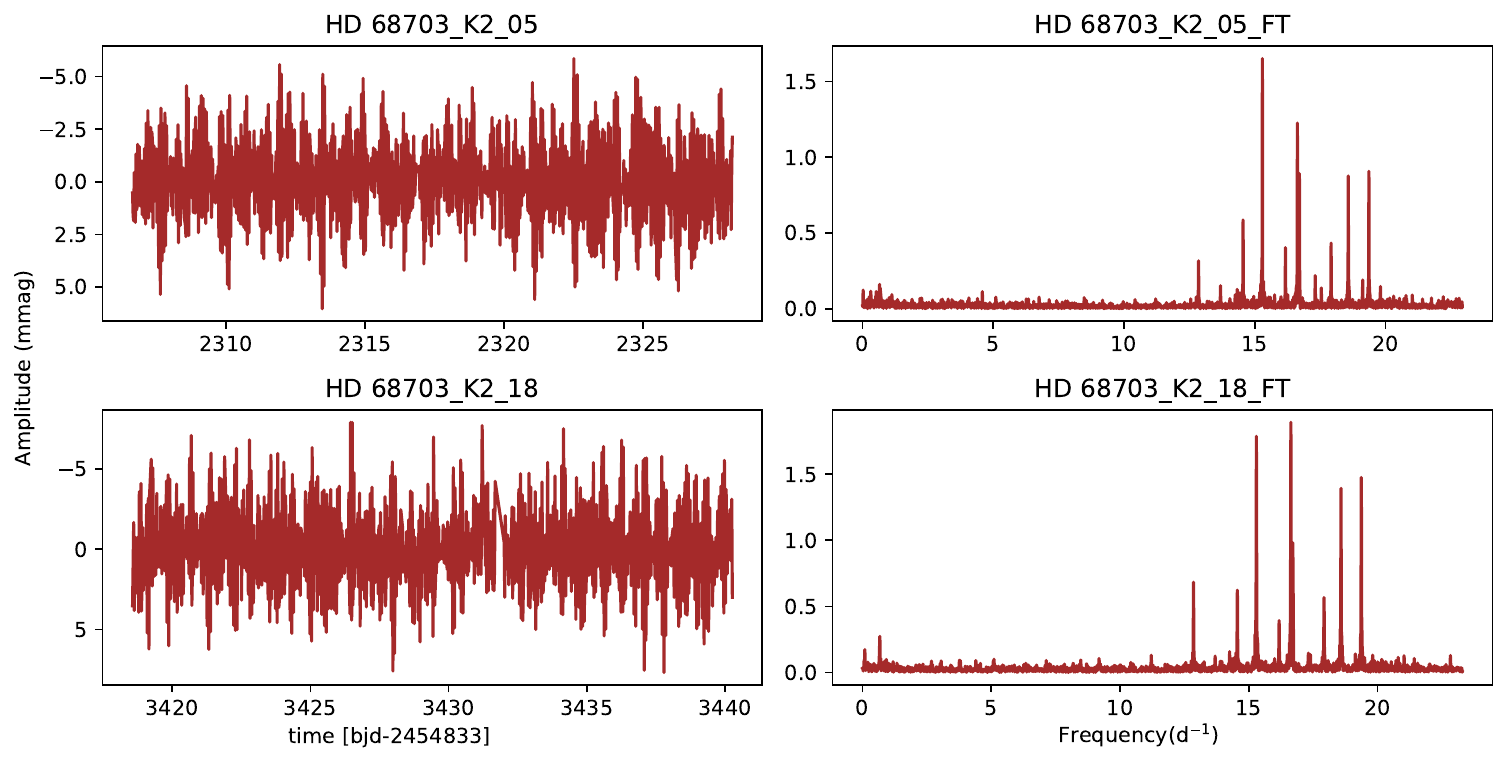}
\caption{
  The light curves (\textit{left panels}) and corresponding frequency spectra (\textit{right panels}) for HD\,68703 obtained with \ktwo\ during Campaign 5 (\textit{top row}) and Campaign 18 (\textit{bottom row}).
}
\label{fresp_k2_68703}
\end{figure*}

\begin{figure*}
\centering
\includegraphics[width=\textwidth]{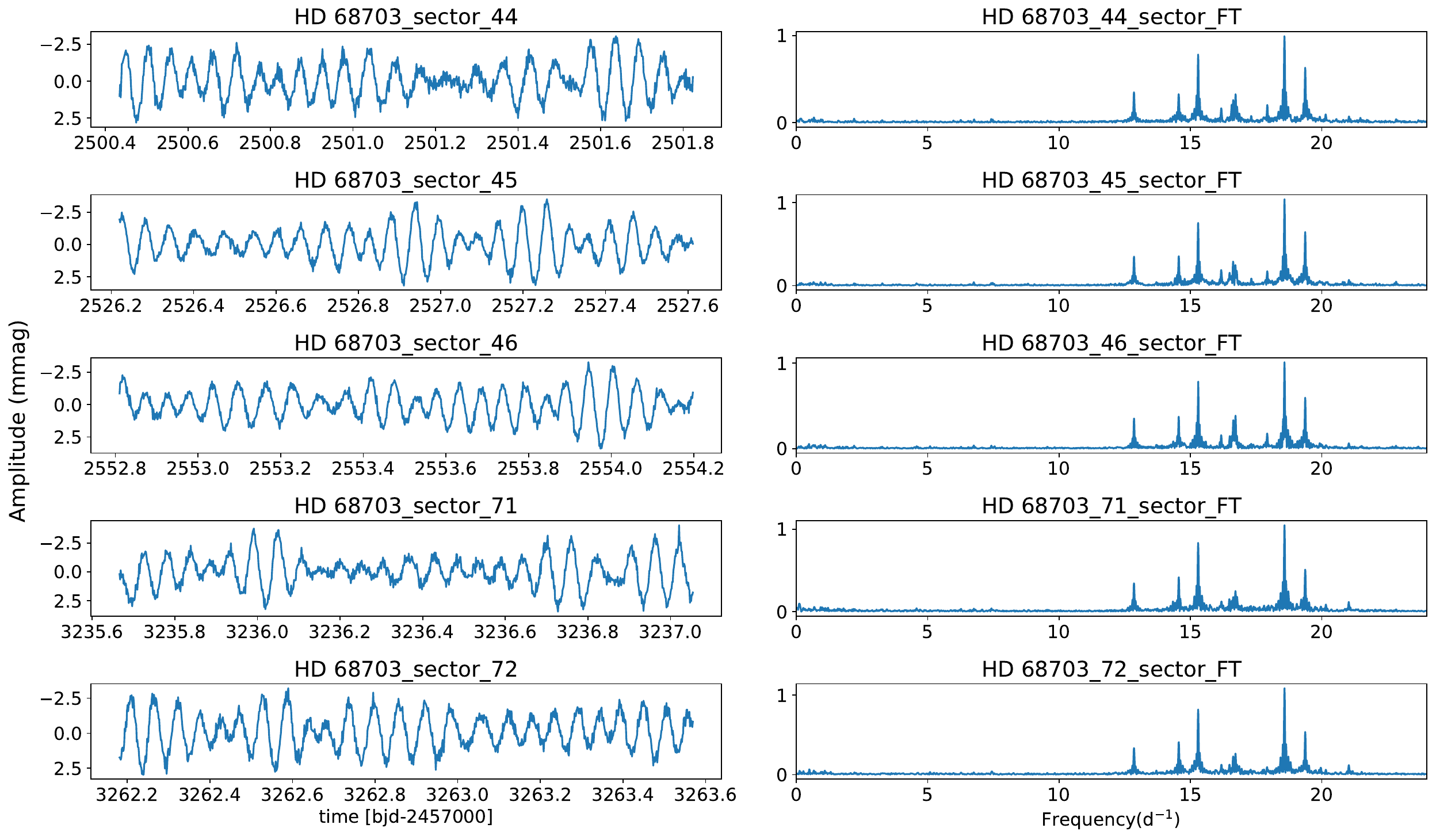}
\caption{The light curves (\textit{left panels}) and corresponding frequency spectra (\textit{right panels}) for HD\,68703 obtained with \tess\ in Sectors 44, 45, 46, 71, and 72 (\textit{from top to bottom row}). We plot only up to 24\,\cd\ since no significant frequency was found beyond this point.}
\label{fresp_tess_68703}
\end{figure*}

\begin{figure*}
\centering
\includegraphics[width=\textwidth]{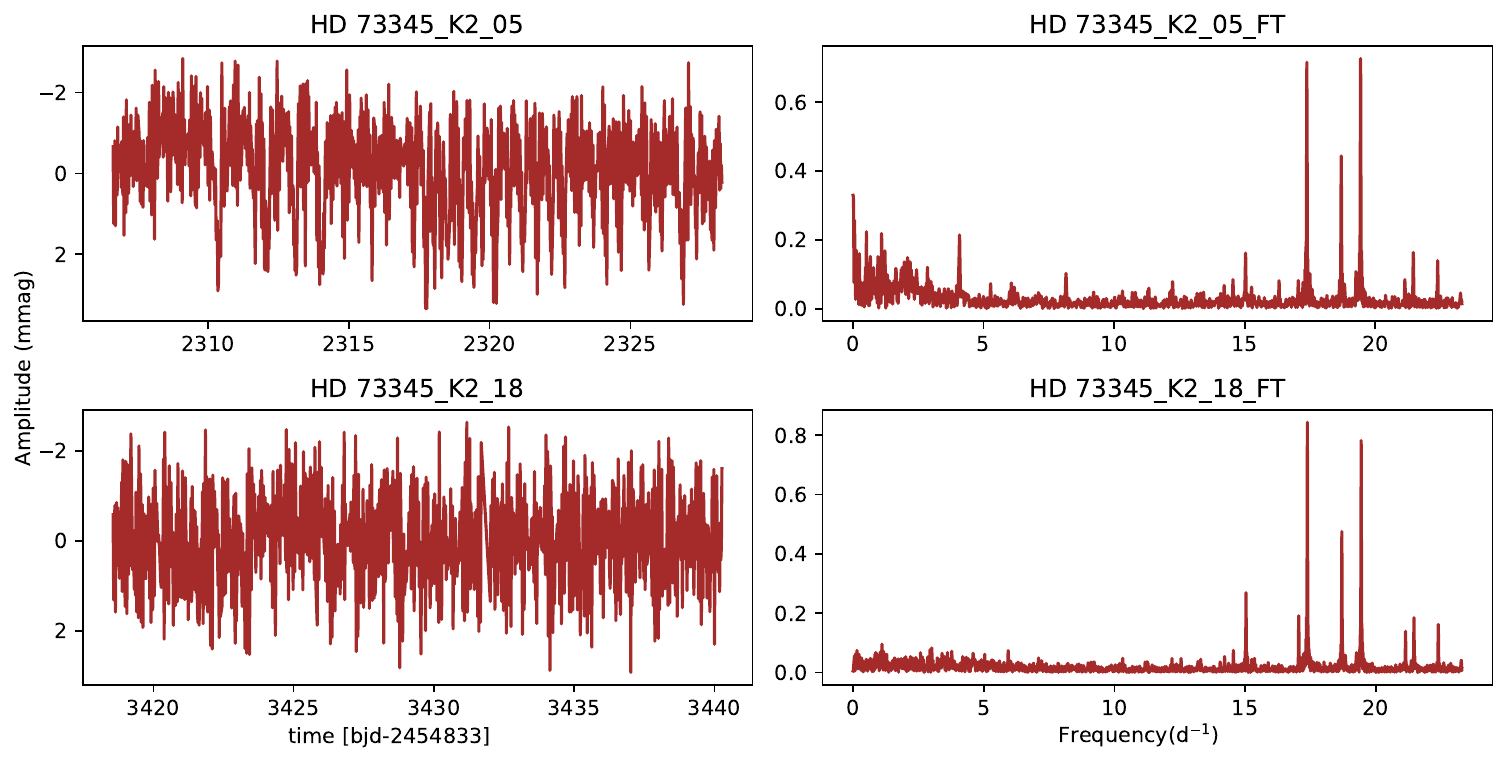}
\caption{
  The light curves (\textit{left panels}) and corresponding frequency spectra (\textit{right panels}) for HD\,73345 obtained with \ktwo\ in Campaign 5 (\textit{top row}) and Campaign 18 (\textit{bottom row}).
}
\label{fresp_k2_73345}
\end{figure*}

\begin{figure*}
\centering
\includegraphics[width=\textwidth]{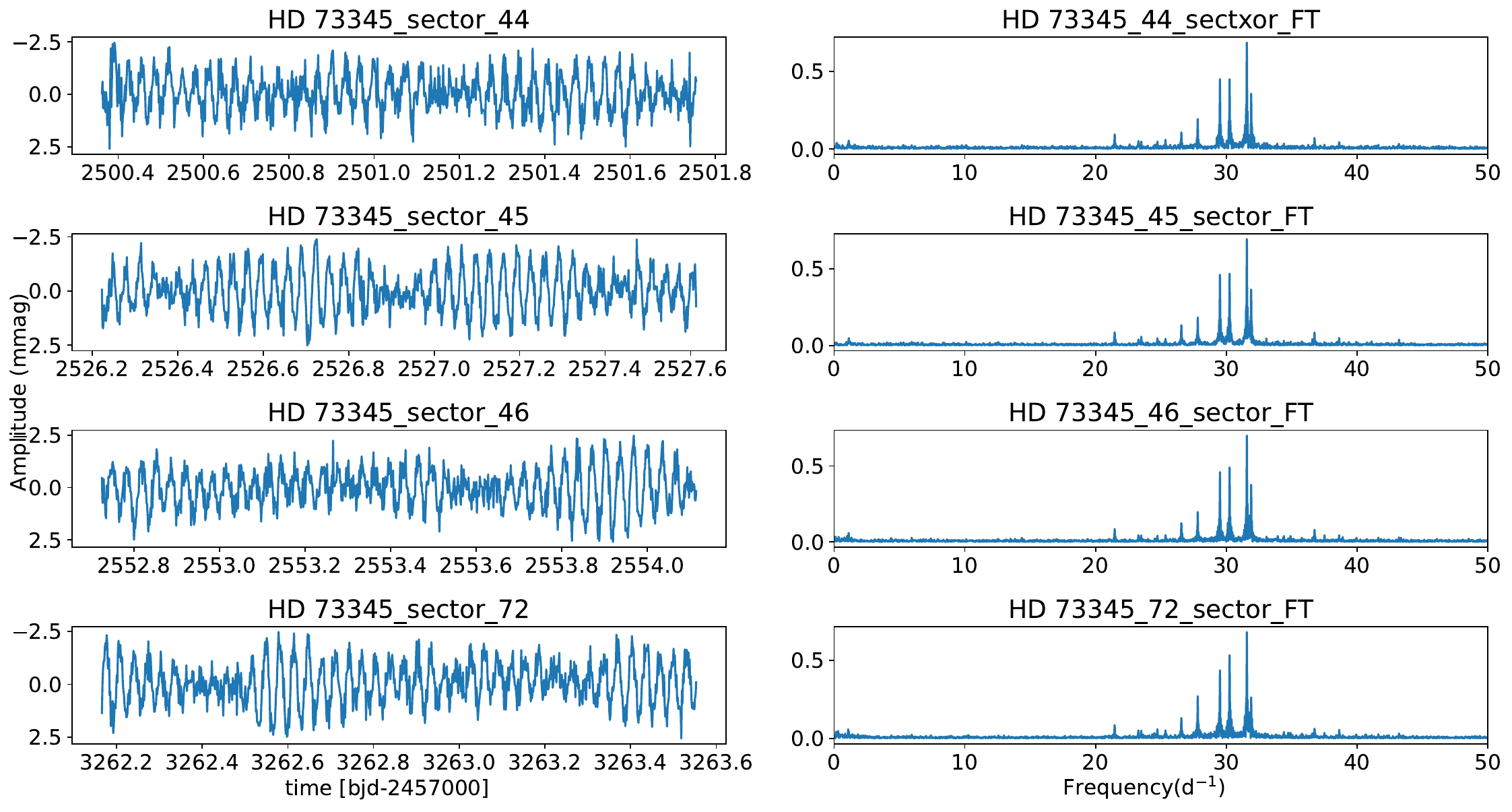}
\caption{
  The light curves (\textit{left panels}) and corresponding frequency spectra (\textit{right panels}) for HD\,73345 obtained with \tess\ in Sectors 44, 45, 46, and 72 (\textit{from top to bottom row}). We plot only up to 50\,\cd\ as no significant frequencies were found beyond this range. The noticeable asymmetry in the light curves may be attributed to the interaction of two closely spaced frequencies.
}
\label{fresp_tess_73345}
\end{figure*}

\newpage

\section{Frequency Solutions from \tess\ and \ktwo\ Time-series Analysis}

\begin{table*}
\centering
\caption{
  The frequencies, corresponding amplitudes, and signal-to-noise ratios (SNRs) for HD\,23734 derived from three consecutive sectors (42, 43, and 44) of \tess\ time-series data. The frequency solution based on the \ktwo\ data from Campaign 4 is also listed. Pulsation constants ($Q$) are provided for the identified radial modes. Only the short-cadence \tess\ data were used in this analysis to avoid aliasing effects and to resolve frequencies beyond the Nyquist (Nyq) limit inherent to \ktwo\ long-cadence data. The column `Comb. Freqs.' indicates the identified radial modes, rotational frequency, and their combinations.
}
\label{tab:tablet1}
\fontsize{6.5}{9.0}\selectfont
\begin{tabular}{ccccccccccc}
\hline\hline
& & \tess &&&&&\ktwo&&\\   
\hline
  & $f$ & Amp  &SNR  & Q-value& Comb. & & $f$ & Amp  &SNR\\
        & (\cd) & ($\pm$0.003 mmag) &  & (days) &Freqs.&& (\cd) & ($\pm$0.1 mmag) &  \\
\hline 
$f_{T1}$&22.90016$\pm$0.00001&1.727&309.8&- &-&$f_{K1}$&22.9001$\pm$0.0007&1.1&22.0\\
$f_{T2}$&25.60789$\pm$0.00003&0.936&175.4&-& -&2Nyq-$f_{K2}$&25.608$\pm$0.001&0.5&10.7\\
$f_{T3}$&16.13569$\pm$0.00004&0.662&126.6&-&- &$f_{K3}$&16.136$\pm$0.002&0.5$\pm$0.1&9.1\\
$f_{T4}$&38.94782$\pm$0.00006&0.446&75.7&0.017 $\pm$ 0.003& F4&-&-&-&-\\
$f_{T5}$&30.78033$\pm$0.00008&0.308&58.4&-&-&-&-&-&- \\
$f_{T6}$&38.79358$\pm$0.00009&0.286&49.0&-&-&-&-&-&-\\
$f_{T7}$&32.93878$\pm$0.00009&0.265&42.1&0.02 $\pm$ 0.004&F3&-&-&-&-\\
$f_{T8}$&42.4168$\pm$0.0001&0.246&44.3&- &-&-&-&-&-\\
$f_{T9}$&22.5310$\pm$0.0001&0.221&40.8&- &-&-&-&-&-\\
$f_{T10}$&25.7210$\pm$0.0001&0.208&39.4&-  &-&-&-&-&-\\
$f_{T11}$&42.9315$\pm$0.0001&0.185&31.1&- &-&-&-&-&-\\
$f_{T12}$&33.0666$\pm$0.0002&0.160&24.3&- &-&-&-&-&-\\
$f_{T13}$&29.8766$\pm$0.0002&0.165&29.6&- &- &-&-&-&-\\
$f_{T14}$&42.2087$\pm$0.0002&0.164&29.5&- & -&-&-&-&-\\
$f_{T15}$&22.0017$\pm$0.0002&0.160&28.8&-&-&-&-&-&-\\
$f_{T16}$&19.2192$\pm$0.0002&0.156&30.8&-& -&-&-&-&-\\
$f_{T17}$&32.2210$\pm$0.0002&0.141&22.2& - &-&-&-&-&-\\
$f_{T18}$&33.9091$\pm$0.0002&0.131&24.4&- &-&-&-&-&-\\
$f_{T19}$&39.5157$\pm$0.0002&0.124&20.6	&-&	-&-&-&-&-\\
$f_{T20}$&41.7995$\pm$0.0002&0.128&23.1&- &-&-&-&-&-\\
$f_{T21}$&49.9145$\pm$0.0002&0.117&18.6&-&-&-&-&-&-\\
$f_{T22}$&26.0984$\pm$0.0002&0.101&18.5& - &-&-&-&-&-\\
$f_{T23}$&44.0720$\pm$0.0003&0.096&16.3&- &-&-&-&-&-\\
$f_{T24}$&22.4092$\pm$0.0003&0.091&16.6&- &-&-&-&-&-\\
$f_{T25}$&49.0387$\pm$0.0003&0.087&12.3&- &-&-&-&-&-\\
$f_{T26}$&37.7358$\pm$0.0003&0.086&15.7&- &-&-&-&-&-\\
$f_{T27}$&36.3113$\pm$0.0003&0.082&12.6&-&-&-&-&-&-\\
$f_{T28}$&27.5403$\pm$0.0003&0.080&15.0&-&-&-&-&-&-\\
$f_{T29}$&42.0281$\pm$0.0003&0.078&14.0&-& - &-&-&-&-\\
$f_{T30}$&31.1892$\pm$0.0003&0.074&13.6&-&-&-&-&-&-\\
$f_{T31}$&54.7124$\pm$0.0003&0.074&12.5&- &-&-&-&-&-\\
$f_{T32}$&34.3933$\pm$0.0004&0.070&11.6&- &-&-&-&-&-\\
$f_{T33}$&24.9947$\pm$0.0004&0.064&11.2&-&-&-&-&-&-\\
$f_{T34}$&46.4479$\pm$0.0004&0.063&9.9&- &-&-&-&-&-\\
$f_{T35}$&40.8229$\pm$0.0004&0.056&10.3&-&-&-&-&-&-\\
$f_{T36}$&43.8136$\pm$0.0004&0.056&9.1&- &-&-&-&-&-\\
$f_{T37}$&52.5559$\pm$0.0005&0.054&9.0&- & -&-&-&-&-\\
$f_{T38}$&29.2187$\pm$0.0004&0.056&11.9&- &-&-&-&-&-\\
$f_{T39}$&29.0548$\pm$0.0005&0.053&11.1&- &-&-&-&-&-\\
$f_{T40}$&49.5389$\pm$0.0005&0.053&8.1&- &-&-&-&-&-\\
$f_{T41}$&28.3355$\pm$0.0005&0.051&10.9&- &-&-&-&-&-\\
$f_{T42}$&21.6257$\pm$0.0005&0.051&9.7&- &-&-&-&-&-\\
$f_{T43}$&18.7759$\pm$0.0005&0.046&9.2& - &-&-&-&-&-\\
$f_{T44}$&33.3351$\pm$0.0005&0.045&7.0&- &-&-&-&-&-\\
$f_{T45}$&54.5173$\pm$0.0006&0.045&7.2&- &-&-&-&-&-\\
$f_{T46}$&18.3873$\pm$0.0006&0.044&8.8&- &-&-&-&-&-\\
$f_{T47}$&30.3839$\pm$0.0006&0.045&8.3& -&-&-&-&-&-\\
$f_{T48}$&29.6660$\pm$0.0006&0.044&8.5&- &-&-&-&-&-\\
$f_{T49}$&31.7962$\pm$0.0006&0.043&7.0&- &-&-&-&-&-\\
$f_{T50}$&46.0459$\pm$0.0006&0.041&6.9& - &-&-&-&-&-\\
$f_{T51}$&10.3633$\pm$0.0006&0.041&9.7&- &-&-&-&-&-\\
$f_{T52}$&37.0475$\pm$0.0006&0.040&7.6&- &-&-&-&-&-\\
$f_{T53}$&41.4372$\pm$0.0006&0.039&7.1&-&-&-&-&-&-\\
$f_{T54}$&36.1861$\pm$0.0006&0.039&5.8&- &-&-&-&-&-\\
$f_{T55}$&56.1051$\pm$0.0006&0.038&6.5&- &-&-&-&-&-\\
$f_{T56}$&50.8394$\pm$0.0006&0.039&7.2&- & -&-&-&-&-\\
$f_{T57}$&37.4168$\pm$0.0006&0.038&6.9&- &-&-&-&-&-\\
$f_{T58}$&21.6767$\pm$0.0006&0.039&7.2&- &-&-&-&-&-\\
$f_{T59}$&47.1537$\pm$0.0007&0.037&6.4&- &-&-&-&-&-\\
$f_{T60}$&12.3188$\pm$0.0007&0.037&7.1& - &-&-&-&-&-\\
$f_{T61}$&18.9915$\pm$0.0007&0.037&7.4&-&-&-&-&-&-\\
$f_{T62}$&52.8368$\pm$0.0007&0.036&5.6&- &-&-&-&-&-\\
$f_{T63}$*&2.3406$\pm$0.0007&0.034&4.8&- &F$_{\mathrm{rot}}$&-&-&-&-\\
$f_{T64}$*&4.660$\pm$0.001&0.018&3.5&- &2F$_{\mathrm{rot}}$&-&-&-&-\\
\hline\hline
\end{tabular}
\begin{minipage}{0.9\linewidth}
F$n$: Frequency of radial mode with order $n$; F$_{\mathrm{rot}}$: Rotational frequency; `*': Frequencies with SNR $< 5.2$, but potentially corresponding to the rotational frequency or its harmonic.
\end{minipage}
\end{table*}

\newpage

\begin{table*}
\centering
\caption{
  Same as Table\,\ref{tab:tablet1}, but for HD\,68703. In this case, we combined the consecutive \tess\ sectors 44, 45, and 46, along with the \ktwo\ light curves from Campaigns 5 and 18. No frequencies beyond the \ktwo\ Nyquist limit were detected in this analysis.
}
\label{tab:tablet2}
\fontsize{7.5}{9.0}\selectfont
\begin{tabular}{ccccccccccc}
\hline
\hline
& \tess &&&&& \ktwo &&\\
\hline
  & $f$ & Amp  & SNR  &Q-value& Comb. & & $f$ & Amp  & SNR \\
        & (\cd) & ($\pm$0.002 mmag) & &(d)& Freqs. &  & (\cd) & ($\pm$0.006 mmag) & \\
\hline 
$f_{T1}$&18.57798$\pm$0.00001&1.032&246.9&0.016$\pm$0.003&F4&$f_{K4}$&18.577989$\pm$0.000006&0.0189&69.9\\
$f_{T2}$&15.29221$\pm$0.00002&0.783&179.3&0.020$\pm$0.003& F3 &$f_{K1}$&15.292443$\pm$0.000004&0.737&119.6\\
$f_{T3}$&19.36695$\pm$0.00002&0.641&145.0 &-& - &$f_{K3}$&19.366866$\pm$0.000006&0.481&63.1\\
$f_{T4}$&14.55409$\pm$0.00003&0.365&106.9&-& -&$f_{K6}$&14.55414$\pm$0.00001&0.268&48.7\\
$f_{T5}$&12.85130$\pm$0.00003&0.358&127.5& -&  F1+3F$_{\mathrm{rot}}$&$f_{K8}$&12.85119$\pm$0.00001&0.207&48.3\\
$f_{T6}$&16.70863$\pm$0.00004&0.326&83.2&-&-&$f_{K5}$&16.709372$\pm$0.000007&0.385&50.9\\
$f_{T7}$&16.63107$\pm$0.00004&0.328&82.9&-&  &$f_{K2}$&16.632078$\pm$0.000004&0.645&84.5\\
$f_{T8}$&17.92033$\pm$0.00007&0.188&47.8&-&- &$f_{K7}$&17.92044$\pm$0.00001&0.213&35.4\\
$f_{T9}$&16.17536$\pm$0.00009&0.139&33.0&-&- &$f_{K9}$&16.17520$\pm$0.00002&0.180&22.4\\
$f_{T10}$&21.0273$\pm$0.0002&0.077&17.1&-&2F3-F1&-&-&-&-\\
$f_{T11}$&16.4889$\pm$0.0002&0.072&17.9&-&- &-&-&-&-\\
$f_{T12}$&17.3132$\pm$0.0002&0.065&18.5&-&-&$f_{K13}$&17.31204$\pm$0.00004&0.078&10.6\\
$f_{T13}$&19.9153$\pm$0.0002&0.050&11.4&-&-&-&-&-&-\\
$f_{T14}$&20.1599$\pm$0.0002&0.052&12.6&-& - &-&-&-&-\\
$f_{T15}$&13.6909$\pm$0.0003&0.046&16.8&-&-&$f_{K18}$&13.69112$\pm$0.00005&0.058&11.4\\
$f_{T16}$&22.8251$\pm$0.0003&0.044&12.7&-&-&-&-&-&-\\
$f_{T17}$&0.9242$\pm$0.0003&0.042&7.0&-&-&-&-&-&- \\
$f_{T18}$&0.6464$\pm$0.0003&0.039&6.3&-&-&-&-&-&- \\
$f_{T19}$&19.9664$\pm$0.0003&0.041&9.4&-& -&-&-&-&-\\
$f_{T20}$&0.4850$\pm$0.0003&0.042&6.7 &-&-&-&-&-&- \\
$f_{T21}$&6.7560$\pm$0.0003&0.040&11.8 &-& - &-&-&-&- \\
$f_{T22}$&2.1989$\pm$0.0003&0.039&9.4& -&2F$_{\mathrm{rot}}$&-&-&-&-\\
$f_{T23}$&7.4194$\pm$0.0003&0.038&12.2 &-&- &-&-&-&-\\
$f_{T24}$&17.3882$\pm$0.0003&0.036&10.0&-&-
&-&-&-&-\\
$f_{T25}$&16.1046$\pm$0.0003&0.039&9.3&-&-&-&-&-&- \\
$f_{T26}$&1.0995$\pm$0.0004&0.033&5.9&-&F$_{\mathrm{rot}}$&-&-&-&-\\
$f_{T27}$&17.8126$\pm$0.0004&0.032&8.6&-& - &-&-&-&- \\
$f_{T28}$&21.4722$\pm$0.0004&0.030&7.1&-&-&-&-&-&- \\
$f_{T29}$&19.8069$\pm$0.0004&0.031&7.3 &-& - &$f_{K17}$&19.80777$\pm$0.00005&0.061&8.4\\
$f_{T30}$&19.6983$\pm$0.0004&0.030&7.3&-&F4+F$_{\mathrm{rot}}$, F3+4F$_{\mathrm{rot}}$ &-&-&-&-\\
$f_{T31}$&3.2738$\pm$0.0005&0.027&7.6& -&3F$_{\mathrm{rot}}$, F4-F3 &-&-&-&-\\
$f_{T32}$&4.5788$\pm$0.0005&0.027&6.8& -&-&-&-&-&-\\
$f_{T33}$&20.5333$\pm$0.0005&0.027&6.0&-&-&-&-&-&-\\
$f_{T34}$&17.5449$\pm$0.0005&0.027&7.2&-&-&-&-&-&- \\
$f_{T35}$&23.6716$\pm$0.0005&0.023&6.6&-& -&-&-&-&-\\
$f_{T36}$&9.5603$\pm$0.0007&0.018&6.5&0.031$\pm$0.005 & F1&-&-&-&-\\
-&-&-&-&-&-&$f_{K15}$&16.58560$\pm$0.00004&0.076&9.8\\
-&-&-&-&-&-&$f_{K16}$&19.13116$\pm$0.00004&0.063&8.9\\
\hline\hline
\end{tabular}
\end{table*}

\newpage

\begin{table*}
\centering
\caption{
  Same as Table~\ref{tab:tablet1}, but for HD\,73345. The frequency solution is given here is on combining the consecutive \tess\ sectors 44, 45, and 46 and \ktwo\ time-series for the Campaigns 5 and 18.
}
\label{tab:tablet3}
\fontsize{7.5}{9.0}\selectfont
\begin{tabular}{ccccccccccc}
\hline
\hline
 & \tess & & & & & \ktwo & & \\
\hline
  & $f$& Amp  &SNR  &Q-value& Comb.  & & $f$& Amp  &SNR \\
 & (\cd)&($\pm$0.003 mmag)&& & Freqs&&&($\pm$0.006 mmag)&\\
\hline 
$f_{T1}$&31.56558$\pm$0.00003&0.694&154.0&- &-&2Nyq-$f_{K1}$&31.56564$\pm$0.00001&0.285&107.2\\
$f_{T2}$&30.24686$\pm$0.00005&0.473&108.4&-& - &2Nyq-$f_{K3}$&30.24688$\pm$0.00001&0.203&77.0\\
$f_{T3}$&29.50692$\pm$0.00005&0.458&112.1&0.014 $\pm$ 0.002&F4&2Nyq-$f_{K2}$&29.50709$\pm$0.00001&0.331&130.0\\
$f_{T4}$&31.89596$\pm$0.00006&0.372&80.8&-& - &2Nyq-$f_{K7}$&31.89610$\pm$0.00004&0.064&22.6\\
$f_{T5}$&27.8107$\pm$0.0001&0.191&42.0&-&-&2Nyq-$f_{K8}$&27.80985$\pm$0.00006&0.048&19.7\\
$f_{T6}$&26.5554$\pm$0.0002&0.123&27.2&-&-&2Nyq-$f_{K6}$&26.55562$\pm$0.00004&0.066&24.8\\
$f_{T7}$&21.4556$\pm$0.0003&0.089&22.8&-&-&$f_{K5}$&21.45574$\pm$0.00003&0.081&32.6\\
$f_{T8}$&36.7432$\pm$0.0003&0.081&20.3&-&-&-&-&-&-\\
$f_{T9}$&1.1119$\pm$0.0004&0.055&8.1& -&F$_{\mathrm{rot}}$&-&-&-&-\\
$f_{T10}$&25.3400$\pm$0.0004&0.051&11.7&-&-&2Nyq-$f_{K13}$&25.34047$\pm$0.00008&0.032&10.5\\
$f_{T11}$&23.4880$\pm$0.0005&0.047&11.3&-&-&$f_{K9}$&23.48595$\pm$0.00008&0.035&12.2\\
$f_{T12}$&24.7246$\pm$0.0005&0.047&10.8&0.017 $\pm$ 0.002&F3&2Nyq-$f_{K10}$&24.72477$\pm$0.00008&0.035&11.8\\
$f_{T13}$&38.6261$\pm$0.0005&0.046&9.8&-&-&-&-&-&-\\
$f_{T14}$&23.2674$\pm$0.0005&0.045&10.8& -&-&-&-&-&-\\
$f_{T15}$&34.3951$\pm$0.0007&0.034&6.4& 0.012 $\pm$ 0.002&F5 &2Nyq-$f_{K12}$&34.39525$\pm$0.00008&0.034&10.7\\
$f_{T16}$&33.0620$\pm$0.0007&0.034&7.0&-&-&-&-&-&-\\
$f_{T17}$&43.2062$\pm$0.0007&0.033&7.6& -&-&-&-&-&-\\
$f_{T18}$&37.5039$\pm$0.0008&0.030&7.2&-&-&-&-&-&-\\
$f_{T19}$&36.5739$\pm$0.0008&0.029&7.4&-&-&-&-&-&-\\
$f_{T20}$&35.7724$\pm$0.0009&0.027&5.8&-&-&-&-&-&-\\
$f_{T21}$&10.095$\pm$0.001&0.022&5.6&-&-&-&-&-&-\\
$f_{T22}$&14.360$\pm$0.001&0.022&5.4& - &3F4-3F3&$f_{K19}$&14.2104$\pm$0.0001&0.022&7.0\\
-&-& -&-&-&-&$f_{K4}$&15.03217$\pm$0.00001&0.210&67.9\\
-&-&-&-&-&-&$f_{K11}$&19.25318$\pm$0.00008&0.035&14.1\\
-&-&-&-& -&-&$f_{K15}$&18.8381$\pm$0.0001&0.025&9.3\\
-&-&-&-&-&-&$f_{K16}$&8.1570$\pm$0.0001&0.023&6.0\\
\hline\hline
\end{tabular}
\end{table*}

\section{LSD Profiles Variations}

\begin{figure*}
\centering
\includegraphics[width=\textwidth]{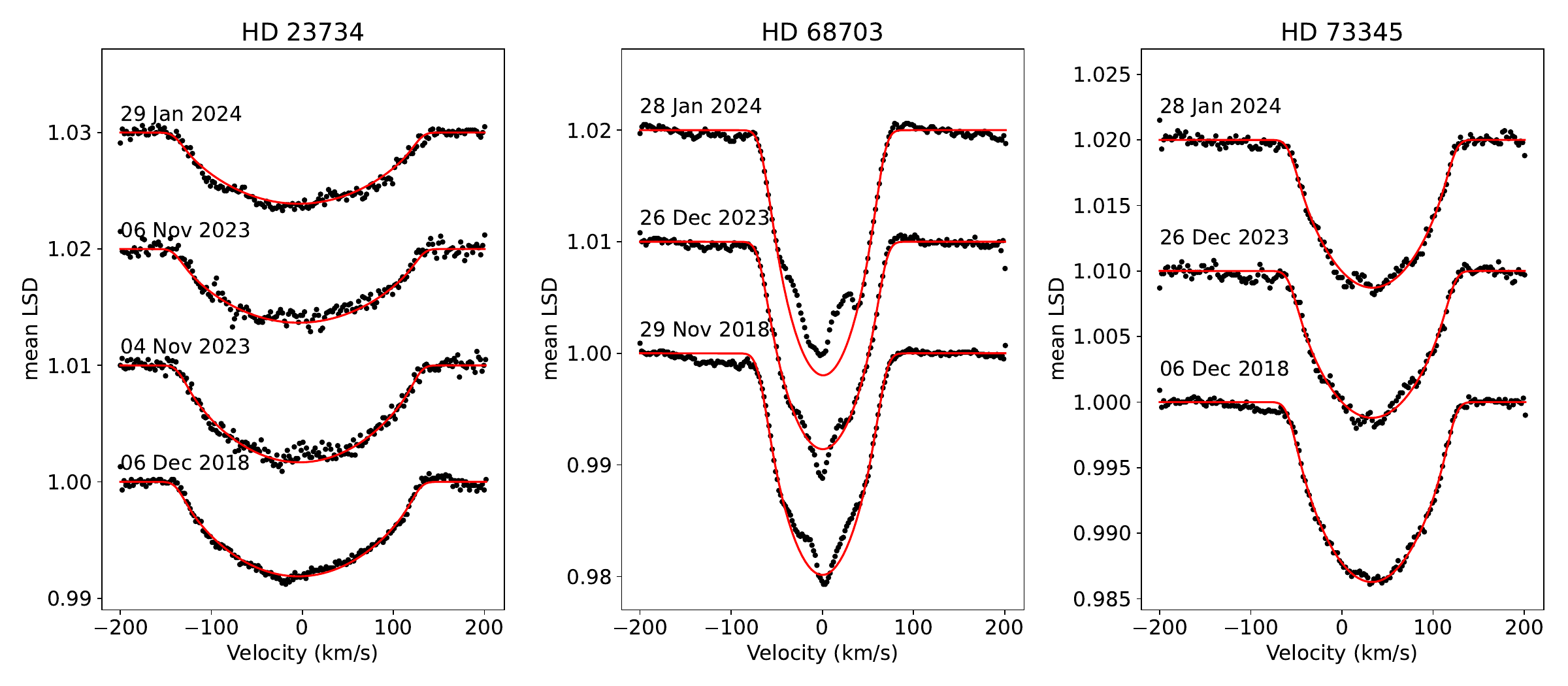}
\caption{
\textit{Left panel:} Mean LSD profiles of HD\,23734 obtained from spectra taken on four different nights. The broadening of the LSD indicates a high rotational velocity, consistent with values listed in Table~\ref{tab:speclog}. \textit{Middle panel} and \textit{Right panel:} LSD profiles of HD\,68703 and HD\,73345, respectively, based on observations from three separate nights. Distinct line profile variations are evident in both cases. In all panels, the red curves represent the best-fit rotational broadening profiles. All thee spectra were acquired using HESP mounted on HCT.
}
\label{fig:LSD}
\end{figure*}

\section{Rotation}

\begin{figure}
\centering
\includegraphics[width=0.5\columnwidth]{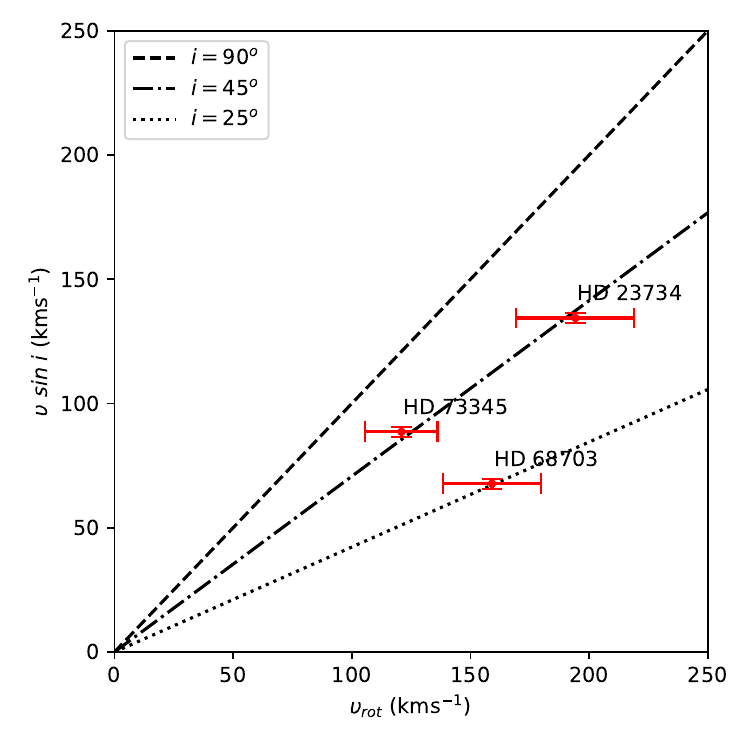}
\caption{Relationship between the equatorial rotational velocity \vrot\ derived from the selected rotational frequency of each star and the corresponding \vsini\ values determined from spectroscopy. The dashed, dash-dotted, and dotted lines corresponds to  inclinations of 90$^{\circ}$, 45$^{\circ}$, and 25$^{\circ}$, respectively.}
\label{fig:v_vsini}
\end{figure} 

\section{Flux Depression and Strength of Ca-K lines}

\begin{figure}
\centering
\includegraphics[width=0.5\columnwidth]{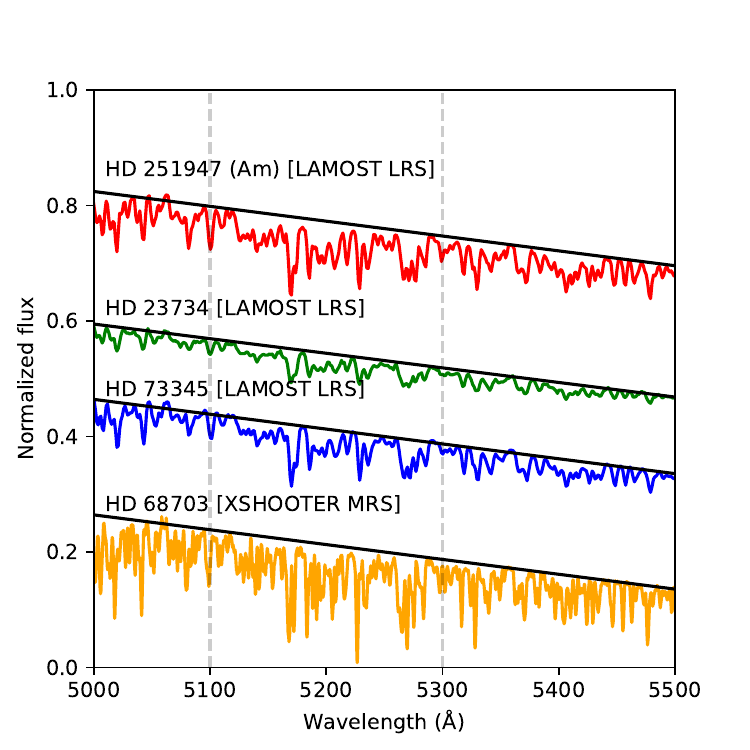}
\caption{Flux calibrated low-resolution LAMOST spectra of HD\,23734 and HD\,73345 (both from DR-10). We could not find a flux-calibrated LAMOST spectrum of HD\,68703 in any of the available data releases, therefore, we took flux calibrated medium-resolution spectra of HD\,68703 obtained by XSHOOTER. For the comparison, flux calibrated LAMOST spectra of a known Am star HD\,251947 (DR-10) is also shown.  The individual spectrum has been vertically offset for clarity. No indication of a flux depression around 5200\, \AA \,(enclosed by two vertical black dashed lines) is observed, providing further evidence of the non-peculiar nature of our sample.}
\label{fig:flux_dep}
\end{figure}

\begin{figure}
\centering
\includegraphics[width=0.5\columnwidth]{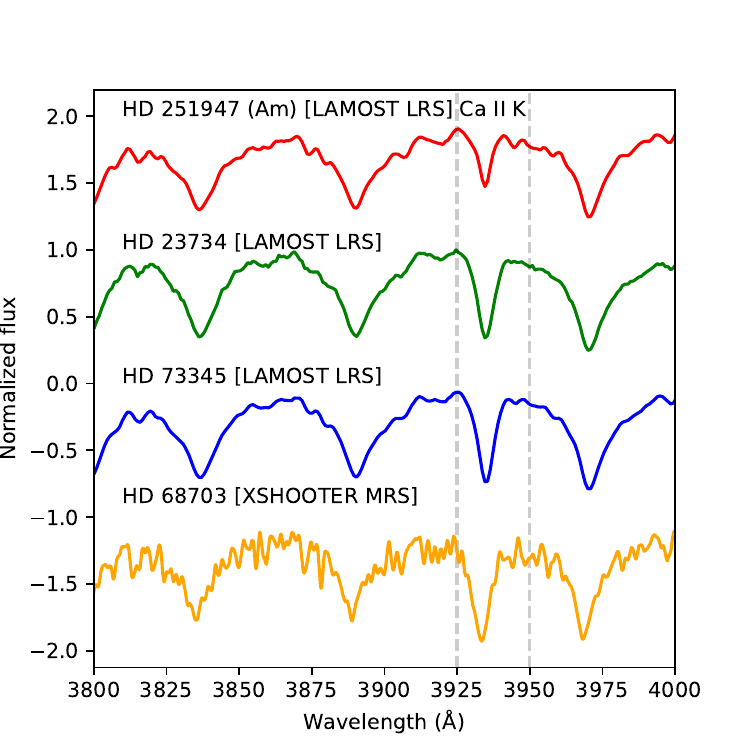}
\caption{ Comparison of the strength of CaII K lines for HD\,23734 and HD\,73345 from LAMOST low resolution spectra with a known Am star HD\,251947 of similar spectral type. Since no LAMOST LRS spectrum is available for HD\,68703, hence, we illustrate the medium-resolution XSHOOTER spectrum instead. From this figure, it is evident that the strength of Ca II K lines is normal in our target stars while the strength of the known Am star, HD\,251947 is weak indicating that the studied stars are chemically normal.}
\label{fig:ca2k}
\end{figure} 


\label{lastpage}
\end{document}